\documentclass[iop]{emulateapj}

\usepackage{rotating}
\usepackage{epsfig}
\usepackage{natbib}
\usepackage{color}
\usepackage{datetime}
\usepackage{wrapfig}

\def\simgt{\lower.5ex\hbox{$\; \buildrel > \over \sim \;$}}
\def\simlt{\lower.5ex\hbox{$\; \buildrel < \over \sim \;$}}

\def\amin{\ifmmode^{\prime}\else$^{\prime}$\fi}
\def\asec{\ifmmode^{\prime\prime}\else$^{\prime\prime}$\fi}

\def\simgt{\lower.5ex\hbox{$\; \buildrel > \over \sim \;$}}
\def\simlt{\lower.5ex\hbox{$\; \buildrel < \over \sim \;$}}

\newcommand\xte{{\it RXTE\/}}

\newcommand\asca{{\it ASCA\/}}

\newcommand\rosat{{\it ROSAT\/}}
\newcommand\chandra{{\it Chandra}}

\newcommand\xmm{{\it XMM-Newton}}

\newcommand\suzaku{{\it Suzaku}}
\newcommand\integral{{\it INTEGRAL}}

\newcommand\swift{{\it Swift\/}}
\newcommand\nustar{\hbox{\it NuSTAR\/}}

\newcommand\eflux{erg\,cm$^{-2}$\,s$^{-1}$}

\def\sgra{Sgr A*}

\def\xrb1743{1E1743.1-2843}
\def\tvcol{TV~Col}
\def\igrsrc{IGR~J17303$-$0601}

\slugcomment{Accepted to The Astrophysical Journal} 

\shorttitle{}
\shortauthors{}

\begin{document}

\title{Evidence for Intermediate Polars as the origin of the Galactic Center Hard X-ray Emission}

\author{Charles~J.~Hailey\altaffilmark{1}, Kaya~Mori\altaffilmark{1}, Kerstin~Perez\altaffilmark{2}, Alicia~M.~Canipe\altaffilmark{1}, Jaesub~Hong\altaffilmark{3}, 
John~A.~Tomsick\altaffilmark{4},  Steven~E.~Boggs\altaffilmark{4}, Finn~E.~Christensen\altaffilmark{5}, William~W.~Craig\altaffilmark{4,6}, 
Francesca~Fornasini\altaffilmark{4}, Jonathan~E.~Grindlay\altaffilmark{3}, Fiona~A.~Harrison\altaffilmark{7}, 
Melania~Nynka\altaffilmark{1, 8}, Farid~Rahoui\altaffilmark{9}, Daniel~Stern\altaffilmark{10}, Shuo~Zhang\altaffilmark{1} and William~W.~Zhang\altaffilmark{11}}

\altaffiltext{1}{Columbia Astrophysics Laboratory, Columbia University, New York, NY 10027, USA; chuckh@astro.columbia.edu}
\altaffiltext{2}{Haverford College, 370 Lancaster Avenue, KINSC L109, Haverford, PA 19041, USA}
\altaffiltext{3}{Harvard-Smithsonian Center for Astrophysics, Cambridge, MA 02138, USA}
\altaffiltext{4}{Space Sciences Laboratory, University of California, Berkeley, CA 94720, USA}
\altaffiltext{5}{DTU Space - National Space Institute, Technical University of Denmark, Elektrovej 327, 2800 Lyngby, Denmark}
\altaffiltext{6}{Lawrence Livermore National Laboratory, Livermore, CA 94550, USA}
\altaffiltext{7}{Cahill Center for Astronomy and Astrophysics, California Institute of Technology, Pasadena, CA 91125, USA}
\altaffiltext{8}{Department of Physics and Astronomy, University of California, Irvine, 4129 Frederick Reines Hall, Irvine, CA 92697}
\altaffiltext{9}{European Southern Observatory, K. Schwarzschild-Strasse 2, D-85748 Garching bei München, Germany} 
\altaffiltext{10}{Jet Propulsion Laboratory, California Institute of Technology, Pasadena, CA 91109, USA}
\altaffiltext{11}{NASA Goddard Space Flight Center, Greenbelt, MD 20771, USA}

%\altaffiltext{2}{Department of Physics, McGill University, Montreal, QC H3A2T8, Canada}
%\altaffiltext{3}{Kavli Institute for Astrophysics and Space Research, Massachusets Institute of Technology, Cambridge, MA 02139, USA}
%\altaffiltext{4}{Space Sciences Laboratory, University of California, Berkeley, CA 94720, USA}
%\altaffiltext{5}{DTU Space - National Space Institute, Technical University of Denmark, Elektrovej 327, 2800 Lyngby, Denmark}
%\altaffiltext{6}{Cahill Center for Astronomy and Astrophysics, California Institute of Technology, Pasadena, CA 91125, USA}
%\altaffiltext{7}{Harvard-Smithsonian Center for Astrophysics, Cambridge, MA 02138, USA}
%\altaffiltext{8}{Department of Astronomy and Astrophysics, The Pennsylvania State University, University Park, PA 16802, USA}
%\altaffiltext{9}{NASA Goddard Space Flight Center, Greenbelt, MD 20771, USA}
%\altaffiltext{10}{Jet Propulsion Laboratory, California Institute of Technology, Pasadena, CA 91109, USA}

\begin{abstract}
Recently, unresolved hard (20-40 keV) X-ray emission has been discovered within the central 10~pc of the Galaxy, possibly indicating a 
large population of intermediate polars (IPs).  
\chandra\ and \xmm\ measurements in the surrounding $\sim$50~pc imply a much lighter 
population of IPs with $\langle M_{\rm WD} \rangle \approx 0.5 M_\odot$.  
Here we use broad-band \nustar\ observations of two IPs: TV Columbae, which has a fairly 
typical but widely varying reported mass of $M_{\rm WD} \approx 0.5$-$1.0 M_\odot$, and  IGR J17303-0601, with a heavy reported mass of $M_{\rm WD} \approx 1.0$-$1.2 M_\odot$.  
We investigate how varying spectral models and observed energy ranges influence  
estimated white dwarf mass.  
Observations of the inner 10~pc can be accounted for by IPs with 
$\langle M_{\rm WD} \rangle \approx 0.9 M_\odot$, consistent with that of the CV population in general, and the X-ray observed field IPs in particular. 
The lower mass derived by \chandra\ and \xmm\ appears to be an artifact of narrow energy band fitting. 
To explain the (unresolved) CHXE by IPs requires an X-ray (2-8~keV) luminosity function (XLF) extending down to at least $5\times10^{31}$~ergs\,s$^{-1}$. 
The CHXE XLF, if extended to the surrounding $\sim$50~pc observed by \chandra\ and \xmm, requires at least $\sim20$-40\% 
of the $\sim$9000 point sources are IPs. 
 If the XLF extends just a factor of a few lower in luminosity, then the vast majority of these sources are IPs. 
This is in contrast to recent observations of the Galactic ridge, where the bulk of the 2-8 keV emission is ascribed to dwarf novae. 
\end{abstract}

%%%%%%%%%%%%%%%%%%%%%%%%%%%%%%%%%%%%%%%%%%%%%%%%%%%%%%%%%%%%

\section{Introduction}
\label{sec:intro}

Recently, {\it Nuclear Spectroscopic Telescope Array} (\nustar) discovered an unresolved central hard X-ray emission (CHXE)  in the inner $\sim$8~pc~$\times \sim$4~pc of the Galaxy~\citep{Perez2015} (KP15).
This emission has a 20-40~keV luminosity of ~$2\times10^{34}$ erg\,s$^{-1}$ and can be described by either thermal bremsstrahlung with $kT >  35$~keV or a power law with photon index $\Gamma \approx 1.2$-1.9 ($N(E) \sim E^{-\Gamma}$). 
KP15 proposes that this emission is due to either stellar origins, such as large populations of intermediate polars 
(IPs), low-mass X-ray binaries, or millisecond pulsars, or diffuse origins, such as cosmic ray outflows from the supermassive black hole Sagittarius A*. 
However, these explanations except the IP interpretation encounter difficulties.
A sufficient low-mass X-ray binary population must have long mean time between outbursts or very faint outburst states to evade Galactic center (GC) monitoring campaigns~\citep{Degenaar2012}. 
The requisite population of several thousand millisecond pulsars would conflict with the dominant IP interpretation of the \chandra\ point source population in the same region~\citep{Muno2004, Pretorius2014}.
A cosmic ray origin has been discussed in the recent work of \citet{Dogiel2015}, but requires fine tuning of the diffusion parameters to reproduce the observed spatial extent. Other possible diffuse sources such as synchrotron radiation from magnetic filaments or low-surface brightness
pulsar wind nebulae are not supported by \chandra\ \citep{Johnson2009} and \nustar\ observations \citep{Mori2015} or supernova birth rates in this region, respectively.

A natural origin of the CHXE could be a population of cataclysmic variables (CVs), in particular IPs.  
CVs are accreting white dwarf (WD) binary systems with short orbital periods ($\la 1$ day) and X-ray emission due to accretion via Roche-lobe overflow from a late-type main sequence companion. 
Magnetic CVs, which have WD magnetic fields strong enough to distort the inner accretion disk, are particularly copious emitters of hard $(> 5$~keV) X-rays.
IPs are a type of magnetic CV that, compared to polars, have longer orbital periods and non-synchronized orbits.
In IPs, infalling material is funneled onto the WD poles along magnetic field lines and heated to temperatures that scale with the WD mass. A standoff shock, with height that adjusts itself to give infalling material time to cool to the photosphere temperature, is the location of the highest temperature material. Below the shock, there exists a column of cooling material, exhibiting a range of temperatures.
Some of the emitted X-rays are viewed through the accretion curtain, requiring an additional absorption term with some partial covering fraction depending on the orientation of the IP with respect to the observer. 
X-rays incident on the WD or pre-shock material may also cause Compton reflection and fluorescent neutral Fe line emission. 
Thus the temperatures and masses obtained for a particular analysis can vary depending on the particular components included in the model.

\chandra\ surveys of the central $2^\circ \times 0.8^\circ$ \citep{Wang2002} and \xmm\ surveys of the central $\sim50$~pc \citep{Heard2013A} of the Galaxy (HW13) have both revealed emission consistent with a large population of magnetic CVs.
\chandra\ has identified $\sim$9,000 point sources~\citep{Muno2009} in the GC, with the faint sources well described by a temperature of $kT \approx 8$~keV~\citep{Muno2004}. 
This is consistent with an IP origin, and is the same temperature observed in \chandra\ studies of the diffuse emission of the central $\sim20$~pc~\citep{MunoDiffuse2004}.  
\xmm\ also observed a diffuse X-ray spectrum described by $kT \approx 8$~keV thermal emission (HW13), and, using accretion shock theory~\citep{Suleimanov2005, Yuasa2012}, inferred that the mass of the WDs is $\sim 0.5 M_\odot$. 
The ratio of 2-10~keV luminosity to stellar mass, $L/M$, increases as it approaches a radius of 4~pc and is on average $\sim4$ times larger than that of the Galactic ridge, implying an IP density $\sim10^2$-$10^3$ times higher than the IP density per integrated stellar mass observed in the solar neighborhood.   

Observations of the Galactic ridge and bulge with \xte\ \citep{Revnivtsev2009}, \suzaku\ \citep{Yuasa2012}, and \integral\ \citep{Krivonos2007} have also established a large population of IPs. 
The hard X-ray emission traces the infrared (IR) distribution associated with stellar populations and is consistent with IP spectra. 
The mean WD mass derived by these observations, also obtained using a model of the cooling post-shock region~\citep{Yuasa2012, Suleimanov2005}, ranges from 0.5 to $0.66M_\odot$, with an IP density more consistent with that found in the solar neighborhood.  

IPs are thus a logical candidate for the CHXE, given their abundance in the solar neighborhood and Galactic center, bulge, and ridge; however, a population of IPs that could account for the CHXE would have peculiar characteristics. 
The CHXE temperature is much higher than that observed by \chandra\ and \xmm\ in nearby regions, and than that observed by \suzaku\ ($kT \sim 15$~keV) in the bulge.  
The mean WD mass obtained by KP15 for the CHXE is $M_{\rm WD} > 0.9 M_\odot$, consistent with the mean mass of field IPs measured by \integral\ and \xmm, $0.86\pm0.07M_\odot$~\citep{Bernardini2012}.  
The heavy IPs detected in the \integral\ IBIS sample have a mean $kT \approx 20$~keV, and $\sim20$\% have $kT \approx 30$-50~keV \citep{Landi2009}. 
However, the CHXE fits to a photon index of $\Gamma = $1.2-1.9, whereas the IBIS sample has a mean photon index of $\Gamma\approx2.7$ and the Galactic ridge and bulge emission fits to $\Gamma\approx2.1$~\citep{Revnivtsev2006, Yuasa2012}.   
Additionally, the CHXE has a spatial distribution that falls much more steeply from the GC than 2-10~keV emission of \chandra\ and \xmm, implying many more massive B-star progenitors in the inner parsecs or a higher efficiency for binary formation or accretion. 

Thus the \nustar\ measurements of the CHXE and the \chandra\ and \xmm\ measurements of the surrounding region imply that two distinct IP populations coexist in the GC: one population constrained to the inner parsecs resembling heavy field IPs and another population filling the surrounding tens of parsecs composed of much lighter IPs. 
At further distance from the GC, the IPs measured in the ridge by  \suzaku, with their much lower $L/M$ and $kT \approx 15$~keV, perhaps imply yet another evolutionary origin (e.g., massive B-stars for the CHXE and later-type stars for the Galactic bulge). 
Alternatively, perhaps the CHXE represents the hot tail of an IP distribution that is detected by \nustar\ in the central parsecs, where a large enough density of these IPs exists to be detected above background.

Further modeling of the CHXE as a heavy IP population, resembling the field IPs studied by \integral, is hampered by the limitations of previous individual IP analyses. 
Firstly, the majority of heavy IP spectral measurements have poorly constrained temperatures/masses, due to either limited statistics or limited observed energy range. 
Secondly, a wide variety of IP spectral models have been applied, introducing systematics in the derived temperatures and masses. 

To address these issues we selected two IPs for detailed study, the moderate-mass TV Columbae (hereafter TV Col) and the relatively heavy \igrsrc\ (hereafter J17303).  
Using the large effective area and energy range of \nustar,  we are able to assess how different spectral models and energy bands cause systematic variations in the mass and temperatures derived.

Our analysis of TV Col and J17303, and its implications for the IP interpretation of the CHXE, proceeds as follows.
In \S\ref{sec:obs}, we introduce these two objects and describe the observations used and data reduction applied. After reviewing various spectral models for the X-ray emission from 
IPs in \S\ref{sec:spec_models}, we present the results of spectral analysis of the two IPs in differing energy bands in \S\ref{sec:spec} and address 
the systematics associated with the WD mass measurements in \S\ref{sec:systematics}. 
We apply these same models to the CHXE spectrum, both using narrow-band (2-10~keV) \xmm\ data and broad-band (2-40~keV) \xmm\ and \nustar\ data, in \S~\ref{sec:CHXE}. 
In \S\ref{sec:discussion} we use these results to demonstrate that a population of IPs with mean WD mass of $M_{\rm WD} \approx 0.87M_\odot$ can consistently account for the CHXE. We will also discuss the implication for IP population in the GC and the Galactic ridge X-ray emission. We summarize our results 
in \S\ref{sec:summary}.
%We offer several possible explanations for how such a population of IPs, with mean WD mass higher than observed in the SDSS survey and apparently requiring a large number of massive B-star progenitors, arises naturally.   

%%%%%%%%%%%%%%%%%%%%%%%%%%%%%%%%%%%%%%%%%%%%%%%%%%%%%%%%%%%%

\section{\nustar\ observations and data reduction}
\label{sec:obs}

{\it NuSTAR} observed TV~Col and J17303 in May and June 2014 for $\sim50$~ks each, as listed in Table~\ref{tab:obs}. 
\nustar\ is the first hard X-ray focusing telescope in orbit, operating from $3$ to $79$ keV. 
It contains two co-aligned optic and detector focal plane modules (FPMA and FPMB), with an angular resolution of $58^{\prime\prime}$ Half Power Diameter ($18$\asec\ full-width at half-maximum, FWHM) and an energy resolution (FWHM) of $\sim400$~eV below $\sim50$~keV and $\sim900$~eV at $68$~keV ~\citep{Harrison2013}. 
Each IP was fully contained on one of the four detector chips throughout the full observation, with the IP placed along the optical axis. 
For the CHXE analysis, we use observations of the Sagittarius A* region in 2012, the three \nustar\ observations listed in Table~\ref{tab:obs} and two archived \xmm\ observations, 0694640301 and 0694641101, reprocessed with SAS v12.0.1. 
These are the same observations used in KP15. 
All \nustar\ data were processed using the \nustar\ {\it Data Analysis Software (NuSTARDAS)} v1.3.1. 

\begin{deluxetable}{lcccc}
\tablecaption{{\it NuSTAR} observations of TV Col, \igrsrc\ and the Galactic Center}
\tablewidth{0pt}
\tablecolumns{4}
\tablehead { \colhead{ObsID}    &   \colhead{Start Date}   &   \colhead{Exposure}   &  \colhead{Target} \\
 \colhead{     }  & \colhead{(UTC)}        & \colhead{(ks)}     }
\startdata
30001002001  & 2012 07 20 &  154 & \sgra \\
30001002003  & 2012 08 04 &  77  & \sgra \\
30001002004  & 2012 10 16 &  50  & \sgra \\
30001020002  & 2014 05 11 &  49  & \tvcol \\
80002013012  & 2014 06 14 &  49  & \igrsrc 
\enddata
\tablecomments {The exposure times listed are corrected for good time intervals.}
\label{tab:obs}
\end{deluxetable}

After filtering for periods of high background during passage through the South Atlantic Anomaly (SAA), we extracted source spectra from a $r=70$\asec\ circular region centered at each source and generated \nustar\ response matrix (rmf) and ancillary response (arf) files using \textbf{nuproducts}.  
Background spectra were extracted from an L-shaped region surrounding, and on the same detector chip as, the source. 
We grouped \nustar\ FPMA and FPMB spectra so that each bin had a source significance of $4\sigma$ ($3\sigma$ for the last bin) above background.  
In joint FPMA and FPMB fits, we include a normalization factor in our fit models to account for the small cross-calibration difference, which is in all cases less than $\sim2.6$\%.
All spectral fitting and flux derivations were performed in XSPEC~\citep{Arnaud1996}, with photoionization cross sections set to those from \citet{Verner1996} and
the abundances for the interstellar absorption set to those from \citet{Wilms2000}. 
Chi-squared statistics were used to assess spectral fitting, and all quoted errors are at 90\% confidence level. 
For both TV Col and J17303, we use energies 3-50~keV, above which detector background is significant.

%%%%%%%%%%%%%%%%%%%%%%%%%%%%%%%%%%%%%%%%%%%%%%%%%%%%%%%%%%%%

\section{Intermediate polar spectral models overview} 
\label{sec:spec_models}

\begin{figure}[t]
\centerline{
\epsfig{figure=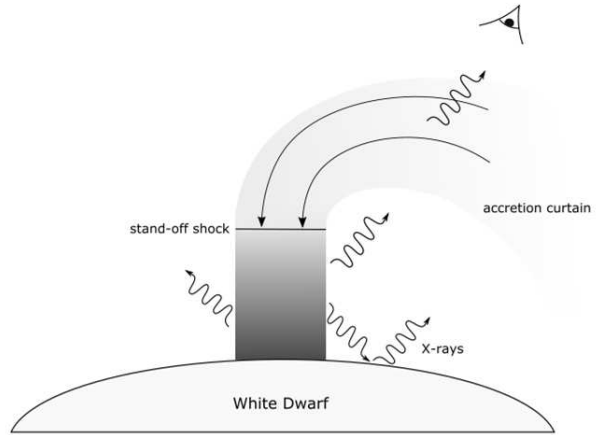,width=1.0\linewidth}                                        
}
\caption{A schematic view of intermediate polar geometry and X-ray emission regions. Primary thermal X-ray photons emitted from the shocked accretion column can be absorbed by the accretion curtain or reflected from the WD surface. Along the post-shock accretion column, plasma density increases toward the WD surface as indicated by the color gradient in the figure. }
\label{fig:ip_cartoon}
\end{figure}

Several spectral models were applied in order to assess the systematics introduced on the derivations of temperature and mass. 
These models include primary X-ray emission from the shocked accretion column (\S\ref{sec:primary_model}) and photon reprocessing via partial 
absorption by the 
accretion curtain and/or reflection from the WD surface (\S\ref{sec:secondary_model}). 
Table~\ref{tab:models} shows a summary of the spectral models, and Figure~\ref{fig:ip_cartoon} illustrates the regions contributing to each emission mechanism.

%\begin{multicols}{2}
%\end{multicols}

\subsection{Primary X-ray emission from shocked accretion column}
\label{sec:primary_model} 

The simplest IP spectral model describes the emission from the accretion column as an optically-thin thermal plasma in collisional ionization equilibrium. 
For this, we use the {\tt APEC} model \citep{Smith2001}, hereafter referred to as the one-temperature or 1T model. 
This is clearly an approximation, since the accretion column in the post-shock region is continuously cooling.
The derived average, or color, temperature is thus a lower limit to the shock temperature.

The most physical IP models utilize 1-dimensional accretion flow models that follow material through the transition shock and down to the WD surface. 
They use either simple emissivity profiles ignoring the soft X-ray line region \citep{Suleimanov2005} or a more detailed soft X-ray emissivity profile \citep{Yuasa2010}. 
We will use the intermediate polar mass (hereafter IPM) model \citep{Suleimanov2005}, which derives a density and temperature profile of the post-shock accretion column and integrates the resultant bremsstrahlung emission as a function of distance from the shock. 
The key parameters of this model are the shock temperature and the WD mass, since the emission depends on the depth of the gravitational potential well, assuming the relation between WD mass and radius of \citet{Nauenberg1972}. 

The disadvantage of the IPM model is that it accounts only for the continuum bremsstrahlung emission (although it does include line cooling) and neglects the ionized line emission, including the prominent Fe lines. 
Acceptable fits to the IPM model can be obtained by ignoring the 5.5-7.5~keV energy bins where H-like and He-like Fe 
lines are found \citep{Ezuka1999}. 
These fits improve with increasing temperature, since the relevant lines become near or fully ionized and thus very weak.  
For instruments with broad-band energy response, such as \nustar, this is sufficient to characterize the WD mass, as shown below. 
This is equally true for the 1T model, where ignoring the line producing region is equivalent to fitting a color temperature to the broad-band bremsstrahlung.   

We also consider a two-temperature (or 2T) model, described by two {\tt APEC} components. 
This will allow comparison with previous analyses of both individual IPs and the GC populations. 
Although not made clear in the literature, this 2T model is strongly motivated as an approximation to the 1-dimensional accretion flow models and 
it is supported by our simulation where we used the \nustar\ responses for consistency with the subsequent spectral analysis. 
We simulated a 3-50 keV \nustar\ spectrum using the IPM model with $M_{\rm WD} = 0.8M_\odot$ and partial covering absorption (described in \S\ref{sec:secondary_model}) with 
$N_{\rm H}=5.0\times10^{23}$~cm$^{-2}$, corresponding to previous measurements of TV Col. 
When fit with a 2T model, a low temperature of $kT_{\rm low} \approx 5$~keV and a high temperature of $kT_{\rm high} \approx 22$~keV are obtained, 
along with the correct column density. 
Similarly, a simulated spectrum using the IPM model with $M_{\rm WD} = 1.0 M_\odot$ and $N_{\rm H} = 2.2  \times 10^{24}$~cm$^{-2}$, corresponding to previous measurements of J17303, is well fit by a 2T model with $kT_{\rm low} \approx 5$~keV and a slightly hotter high temperature of $kT_{\rm high} \approx 30$~keV. In both cases, deviations between the simulated IPM spectra and the best-fit 2T models are mostly within 5\% while they increase up 
to $\sim10$\% only at $E\sim50$ keV.  
Depending on the details of instrument response and whether Fe lines are included in the spectra, the low-temperature component varies from 
$kT_{\rm low} \approx $ 3~keV to 8~keV for $M_{\rm WD} \sim0.5$-$1.0 M_\odot$ IPs. 
This 2T phenomenological model is further supported on theoretical grounds. 
The high temperature component from the immediate vicinity of the transition shock dominates hard X-ray emission thus it closely tracks the shock temperature 
which is proportional to the WD mass, while the low temperature component originates from near the bottom of the accretion column. 
A plot of the thermal bremsstrahlung emissivity $\sim \rho^2  T^{1/2}$ (where $\rho$ and $T$ are density and temperature, respectively) from \citet{Yuasa2010} shows a peak in emissivity at $kT\la10$~keV for massive IPs, which can account for the low-temperature component.  

Another primary X-ray emission model used by previous studies associates the low-temperature component of the 2T model with an ionization temperature through emission line ratio fitting. For most IPs, a ratio of He-like and H-like Fe emission lines at 6.7 and 6.9~keV can be used as a temperature diagnostics for thermal plasma in ionization equilibrium \citep{Mewe1985}. 
Utilizing a large sample of IPs, \asca\ obtained ionization temperatures of $\sim$7-13~keV, with a mean of $kT_{\rm ion} = 8.1$~keV \citep{Ezuka1999}.  
This temperature correlates well with the lower of the temperatures in the two-temperature fits, and corresponds to the region of peak line cooling. 
The \asca\ high temperatures were crudely estimated to be $\sim$15-30~keV, far above the \asca\ energy band, but consistent with the continuum temperatures observed in joint \xmm\ and \integral\ fits to an ensemble of IPs, $\sim$15-40 keV \citep{Landi2009}. 
In \nustar\ broad-band data, however, we will see that small changes to the high-temperature component can cause large changes to the parameters obtained for the low-temperature component. 

If we fit the \nustar\ data in the narrow energy band of 3-10~keV to a single \emph{effective} ionization temperature, this largely decouples the spectral fitting in the soft X-ray band from large variations introduced by an additional temperature associated with the hard X-ray continuum.  
Moreover, exploiting this approach will afford some insight into previous interpretation of \asca\ IP spectral fitting and to the \xmm\ fits to the CHXE, discussed in \S~\ref{sec:CHXE}.  
We call this an effective ionization temperature because it is a temperature determined largely from line ratios, but since the accretion column is continually cooling towards the surface, the temperature is only defined in a mean sense. 
However as discussed above, the emissivity will have a maximum at $kT \la 10$~keV, so that we expect a single temperature fit should adequately represent the line structure. 
Then using a model relating effective ionization temperature and WD mass \citep{Aizu1973}, we should recover the WD mass obtained from our broad-band energy fits. 

\subsection{X-ray reprocessing by accretion curtain and white dwarf surface}
\label{sec:secondary_model}

As mentioned in \S\ref{sec:intro}, phase-resolved spectroscopy provides evidence for an accretion curtain.
In phase-averaged observations, this manifests itself as an absorption component that is local to the WD and covers part of the emitting region, and is in addition to the usual interstellar absorption.  
The exact location of the curtain is unclear, but given that it is neutral or warm-ionized, a likely location is above the transition shock. 
Most of the models we present below include a partial covering absorption term, using the {\tt pcfabs} model in XSPEC with neutral hydrogen column density $N^{\rm pc}_{\rm H}$ and covering fraction $f^{\rm pc}$, as a necessary component to obtain good fits. 

The most physically plausible model should also account for X-ray reflection off the WD surface. 
X-rays from the cooling accretion column will reflect from this surface, leading to both Compton reflection of the hard X-rays and neutral or nearly neutral Fe-K fluorescence and absorption features. 
While reflection-like features will be produced by cold or warm absorbing material in the pre-shock region, 
such material may or may not be present; but reflection from the surface should always be present, since the transition shock is located close 
enough to the surface that the solid angle for interception of downward radiation is approximately unity.   
Neutral Fe features with typical equivalent widths of $\approx 100$-200~eV are prominent in many IPs \citep{Ezuka1999}.  
Evidence of reflection from WD surfaces has been tentatively claimed using \asca\ data 
\citep{Ezuka1999} and definitively seen using \nustar\ data \citep{Mukai2015}. 

To model X-ray reflection from the WD surface, we use the convolution model {\tt reflect} \citep{Magdziarz1995} for reprocessing the primary X-ray continuum and a Gaussian line at 6.4~keV for the neutral Fe line.
TV Col is a particularly attractive candidate for reflection modeling because the observation angle between its rotation axis and the earth has been well constrained by previous observations. 
Unfortunately, the partial covering model we apply can also account for some of the X-ray reflection effects, as pointed out by \citet{Yuasa2010}. 
This will lead to considerable parameter degeneracy between the two models.  
%This is an ad hoc procedure; the neutral Fe line is not fit self-consistently in any of these models, including the reflection model.
We note that \citet{Cropper1998} found that the inclusion of a reflection component along with the partial covering model did not significantly change 
the derived WD mass. 
However, given the combination of good statistics and broad-band coverage of \nustar, additional investigation seems warranted.

\begin{deluxetable*}{lll}
\tablecaption{Ten spectral models used for fitting \nustar\ spectra of the two IPs}
\tablewidth{0pt}
\tablecolumns{3}
\tablehead { \colhead{Model name}  & \colhead{Description} & \colhead{XSPEC model \tablenotemark{a} }}
\startdata
1T     & Single temperature optically-thin thermal  & {\tt apec+gauss} \\
IPM   & IP mass model & {\tt ipm} \\
1T-PC   & Partially absorbed 1T optically-thin thermal  & {\tt pcfabs*apec+gauss}   \\
IPM-PC &  Partially absorbed IP mass model &  {\tt pcfabs*ipm} \\
1T-RFL  & 1T optically-thin thermal with reflection  &  {\tt reflect*apec+gauss} \\
IPM-RFL  & IP mass model with reflection & {\tt reflect*ipm} \\
IPM-PC-RFL & Partially absorbed IP mass model with reflection & {\tt pcfabs*reflect*ipm} \\
1T-IPM & Single temperature optically-thin thermal plus IP mass model & {\tt apec+ipm} \\
2T-PC  & Partially absorbed 2T optically-thin thermal & {\tt pcfabs*(apec+apec+gauss) or apec+pcfabs*apec+gauss} \\
%2T-PC-RFL & Partially absorbed 2T optically-thin thermal with reflection & {\tt pcfabs*reflect*(apec+apec+gauss)} \\
3T-PC & Partially absorbed 3T optically-thin thermal & {\tt apec+pcfabs*(apec+apec+gauss)}
\enddata
\tablecomments {2T models are fit with partial absorption of both APECs and of only one APEC. Thermal models are fit with and without iron lines.} 
\tablenotetext{a}{All models are multiplied by {\tt tbabs} to account for Galactic neutral hydrogen absorption. IPM model was implemented as an additive table model {\tt polarmodel.fits}. }
\label{tab:models}
\end{deluxetable*}

%%%%%%%%%%%%%%%%%%%%%%%%%%%%%%%%%%%%%%%%%%%%%%%%%%%%%%%%%%%%

\section{Spectral analysis of TV~Col and \igrsrc}
\label{sec:spec}

Previous studies of field IPs often exhibited large discrepancies in the continuum temperatures and WD masses due to using different
energy bands and spectral models. In order to compare our results directly with the previous temperature/mass measurements and assess their
systematic errors (\S\ref{sec:systematics}), we fit a variety of spectral models shown in Table \ref{tab:models},
including the simplest 1T model, to different combinations of the primary emission (1T, 2T, IPM) and reprocessing models (PC, RFL), to \nustar\ data of TV~Col and J17303. We also determine the ionization temperature by fitting a 1T model with partial covering in the 3-10 keV band 
and our results are discused in \S\ref{sec:limited_band} in comparison with the \asca\ results by \citet{Ezuka1999}.  
We apply two different approaches to fit the IPM model to measure the WD mass. 
First, we fit the full 3-50~keV energy band, excluding Fe lines at 5.5-7.5~keV, with an IPM model with partial-covering (PC) and/or
X-ray reflection from the WD surface (RFL). Second, we fit the \nustar\ data above $E\sim$10-15~keV with an uncovered IPM model. 
Further discussions on the WD 
mass measurements can be found in \S\ref{sec:wd_mass}. We present figures and tables associated with our fitting results in the Appendix.

\subsection{TV Col}
\label{sec:tvcol_results}

TV Col was first discovered in X-rays by {\it Ariel 5}~\citep{Cooke1978}, and subsequently observed by \rosat\ \citep{Vrtilek1996}, \xte\ \citep{Rana2004}, \swift\ 
\citep{Brunschweiger2009} and \suzaku\ \citep{Yuasa2010}.
It is located nearby ($d = 368$~pc), with an associated $V\sim13$ late-type dwarf companion~\citep{McArthur2001} and fairly typical spin and orbital periods 
(spin period $P_{\rm spin} \approx1.9$~ks, orbital period $P_{\rm orb} \approx 5.5$ hours, and $P_{\rm spin}/P_{\rm orbit} \approx 0.1$).
It exhibits both positive and negative superhumps~\citep{Retter2003}.
It has shown optical outbursts of approximately two orders of magnitude, which have been associated with increased mass transfer from the 
companion~\citep{Hellier1993}. Analysis of line emission during the outbursts suggests it comes from impact of the accretion stream with the outer parts 
of a precessing accretion disk.
The observed orbit phase-resolved spectroscopy and energy-dependent power spectral analysis have been interpreted to suggest primarily disk-fed accretion or 
accretion from both the disk and stream \citep{Rana2004}, which is well described by the accretion curtain model with the observer at a modest  inclination 
angle \citep{Rosen1988}. A range of temperatures, $kT \approx $20-40~keV, have been deduced for TV Col, depending on the particular form of the spectral model applied.
While broad, this temperature range is consistent with the majority of IPs observed by \integral\ surveys~\citep{Landi2009}.

\subsubsection{Thermal plasma model fitting}

% 1T model 
The 1T models without partial covering gives a reduced $\chi^2 = 1.33$ with $kT= 20.7\pm0.4$~keV (top left panel of Figure~\ref{fig:tvcol}).
However, phase-resolved studies suggest that an accretion curtain is present, and thus a partial covering model should be used \citep{Rana2004}. 
The 1T model with partial covering (1T-PC) gives a lower temperature of $kT = 15.6\pm0.4$~keV and 
$\chi^2_\nu \approx 1.2$ (the top right panel of Figure~\ref{fig:tvcol}). The 1T model with reflection (1T-RFL) gives a similar temperature of $kT = 16.7^{+0.7}_{-1.9}$~keV and $\chi^2_\nu \approx 1.2$.\footnote[1]{Note that the observation angle for the reflection model is frozen to the value, $\cos(\theta) = 0.3$ corresponding to $\sim70^\circ$ inclination angle, measured previously using phase-resolved data \citep{Rana2004}.} 
We additionally fit the narrow-band \nustar\ 3-10~keV data with a 1T model with partial covering, to obtain an effective ionization temperature of $kT_{\rm ion} = 9.8 \pm 0.9$~keV.

The temperature derived by \swift/BAT \citet{Brunschweiger2009}, using a 1T bremsstrahlung model without partial covering, agrees with the temperature we derive using the 1T model ignoring
the line emission region
(in which case the APEC closely approximates a bremsstrahlung). Excluding the 5.5-7.5~keV line emission region, as was done by \citet{Ezuka1999}, leads to
systematically higher temperatures and better fits for one temperature models, regardless of whether the model has a partial covering or reflection.
This is readily understandable, since a high temperature will fit the hard X-ray emission from near the transition shock, but will over-ionize the line emission region,
which requires a lower temperature in the cooling accretion column.

% 2T model 
Fitting an uncoverd 2T model and 2T model with a partial covering over both temperature components yield $\chi^2_\nu = 1.2$ and 1.1, respectively 
(two middle panels in Figure~\ref{fig:tvcol}).  
A partial covering over both temperature components is a physically plausible representation of an accretion curtain where 
the neutral material may be above the transition shock. 
The best fit value of $kT_{\rm low} \approx $2-3~keV corresponds closely to the temperature at the base of the accretion column for IPs 
simulated with $M_{\rm WD} = 0.8M_\odot$ and it is consistent with theoretical expectations discussed in \S\ref{sec:primary_model}. 
The partial covering model gives hard X-ray temperatures of $kT_{\rm high} \approx 20$~keV.  
The best-fit $kT_{\rm high}$ and $kT_{\rm low}$ remain consistent whether only one or both temperature components are partially covered.

% Explanation for higher ISM nH values 
The best-fit interstellar absorption column densities are higher than the value 
obtained by \citet{Rana2004}: $N^{f}_{\rm H} = 2.1^{+0.2}_{-0.1}\times10^{20}$~cm$^{-2}$.   
The low column density of \citet{Rana2004} is based on jointly fitting  \rosat\ and {\it RXTE} spectra of TV~Col from 0.1~keV to 20~keV and it is consistent with 
the source distance of $368^{+17}_{-15}$~pc \citep{McArthur2001}. 
The discrepancy is due to the lack of low energy data below 3~keV for our \nustar\ analysis as 
\citet{Yuasa2010} also measured a similarly high column density ($N^{f}_{\rm H} = 3.4^{+1.3}_{-2.6}\times10^{22}$~cm$^{-2}$) by fitting the 3-50 keV \suzaku\ spectra of TV~Col.
In addition, $N^{f}_{\rm H}$ is poorly constrained for some models with partial covering largely due to the degeneracy between 
$N^{f}_{\rm H}$ and $N^{pc}_{\rm H}$.  
To investigate the parameter sensitivity to $N^{f}_{\rm H}$, we refitted the \nustar\ spectra of TV~Col by fixing $N^{f}_{\rm H}$ to the value of \citet{Rana2004} for all the models presented in 
Table~\ref{tab:tvcol_fit}. Both the  
temperature and the WD mass of TV~Col remained unchanged within the errors. Therefore, we conclude that our main results and interpretations based on the temperatures and WD masses are not affected 
by the systematic 
errors associated with the interstellar absorption.

Sub-solar abundances and equivalent widths in the $\sim$100-200 eV range, consistent with all the models shown, are observed for most field IPs. 
The  {$\tt rel_{refl}$} parameter measures the angle-averaged fraction of downward radiation intercepted by the WD surface, and the high values seen here are consistent with the theoretical expectation that the transition shock is extremely close to the surface \citep{Yuasa2010}. 
The statistical uncertainties in the derived temperatures and WD masses are small. 
The systematic error, due to different applied models, has not been fully considered by previous observers. 
Averaging the models here yields a mean temperature of $kT = 21$~keV, with a systematic uncertainty of $\sim3.5$~keV estimated as the spread in 
$kT$ obtained from different 1T models. 
The 2T model produces substantially lower values of partial covering column densities compared to the 1T model, and the reflection model does not require any partial covering component to produce acceptable fits. 

% 
%We additionally fit the narrow-band \nustar\ 3-10~keV data for \tvcol\ with a 1T model with partial covering, to obtain an effective ionization temperature of $kT_{\rm ion} = 9.8 \pm 0.9$~keV (see Table~\ref{tab:tvcol_fit}).
%For the 2T model fits discussed above, where  $kT_{\rm low} \ll kT_{\rm high}$, the high-temperature component varies little with the particulars of the model due to the broad energy band of \nustar, and thus 
%can be reliably interpreted as the continuum color temperature.  
%As discussed in \S\ref{sec:secondary_model} and shown above, however, small changes in the high temperature, when extrapolated to the low-energy band, can change the soft X-ray continuum substantially. 
%We thus use the narrow energy band of 3-10~keV and a partially covered 1T model to extract our effective ionization temperature.

\subsubsection{IP mass model fitting}

% IPM model fit 
Using the full 3-50~keV band, the IPM model with partial covering (IPM-PC) and the IPM model with reflection (IPM-RFL) yield 
consistent WD masses of $M_{\rm WD} = 0.80^{+0.01}_{-0.02} M_\odot$
and $M_{\rm WD} = 0.78\pm0.03 M_\odot$, respectively. 
Even with the much higher sensitivity and counting statistics of the \nustar\ observations, there is no
definitive evidence that a reflection model is preferred.
To the contrary, the partial covering model seems
almost completely degenerate with the reflection model, producing comparable fit quality, abundances,
temperatures and masses.
Thus we do not report results for a model with both reflection and partial covering.
The mass in our IPM-PC model are consistent with the \suzaku\ measurements ($M_{\rm WD} = 0.91^{+0.14}_{-0.10} M_\odot$) 
of \citet{Yuasa2010} using a 1D accretion flow model (called IP-PSR) including both the 
continuum and emission lines. 
Thus \nustar\ broad-band energy measurement in which line emission is ignored produce WD masses that are consistent with those obtained by 
various observatories. 

We also fit the 15-50 keV \nustar\ data with an uncovered IPM model and it yields a 
good $\chi_\nu^2 = 0.96$ since the effect of partial covering is negligble at such high energies.  
The best-fit WD mass of $M_{\rm WD} = 0.77\pm0.03 M_\odot$ from the $E>15$~keV fit is in good agreement with the 3-50~keV band IPM-PC fit results 
($M_{\rm WD} = 0.80^{+0.01}_{-0.02} M_\odot$ and $\chi_\nu^2 = 1.14$) as well as the masses derived by \swift/BAT ($M_{\rm WD} = 0.78\pm0.06 M_\odot$, 
14-195~keV, \citet{Brunschweiger2009}) and by \suzaku\ ($M_{\rm WD} = 0.87^{+0.53}_{-0.18} M_\odot$, 15-40~keV, \citet{Yuasa2013})  
using an IPM model without partial covering.

%%%%%%%%%%%%%%%%%%%%%%%%%%%%%%%%%%%%%%%%%%%%%%%%%%%%%%%%%%%%%%%%%%%%%%%%%%%%%

\subsection{IGR~J17303$-$0601}
\label{sec:igr_results}

Unlike TV Col, J17303 has extreme properties for an IP with spin period $P_{\rm spin} = 128$~s, orbital period $P_{\rm orb} \approx 924$~min and 
$P_{\rm spin}/P_{\rm orbit} \approx 2\times10^{-3}$.
It has the second shortest spin to orbital period ratio of any known IP~\citep{Scaringi2010}.
The multiple emission components can be modeled by several optically thin thermal plasmas, with evidence for several absorbing components, including a 
warm (O~VII) absorber that suggests possible photo-ionization of pre-shock material.
The \nustar\ hard X-ray spectrum, discussed below, is also peculiar, showing evidence of reprocessing from one or more sites around the WD.
Observations have consistently yielded a temperature of $kT \approx 60$~keV, placing J17303 among the hottest in the \suzaku\ survey.
Due to its brightness and extremely hard X-ray spectrum, J17303 is the optimal IP to compare the \nustar\ measurement of temperature and WD mass
with the \swift\ \citep{Brunschweiger2009} and \integral\ \citep{Landi2009} results that extends to higher energy than \nustar.

\subsubsection{Thermal plasma model fitting}
% 1T model fit 
Unlike in the case of TV Col, a 1T model without partial covering or reflection shows large residuals in the 10-40~keV energy band 
and a high $\chi_\nu^2 = 1.90$ (upper left panel Figure~\ref{fig:igr} in the Appendix).  
Such a high-energy ``hump" can result from strong reflection from the WD surface, but could also 
be produced by strong absorption in an accretion curtain.  
A 1T model with partial covering (1T-PC) gives an excellent $\chi_\nu^2 =1.05$, even including the Fe line-emitting 5.5-7.5~keV region (upper right panel 
in Figure~\ref{fig:igr}), and yields $kT =  26\pm1$~keV, which is typical of many individual IPs even
though J17303 is an outlier in period-spin space. 
On the other hand, the narrow-band \nustar\ 3-10~keV data fit with a 1T model with partial covering yields an effective ionization temperature of 
$kT_{\rm ion} = 10.2_{-1.1}^{+3.5}$~keV.
For the reflection model, the observation angle 
was left as a free parameter, since there is no independent constraint.
The reflection 1T model yields $\chi_\nu^2 =1.03$ with $kT =  41^{+3}_{-2}$~keV, 
However, the best-fit value of ${\tt rel_{refl}}$, which roughly characterizes the fraction of the downward incident radiation reflected by the 
WD surface, is greater than unity.  
To constrain this unphysical value, the results in Table~\ref{tab:igr_fit} are shown for a value of  {$\tt rel_{refl}$} frozen to one.

% 2T model fit 
Similar to the 1T model fits, a partial covering model component is required to fit a 2T model to the 3-50 \nustar\ spectra. 
Fitting a 2T model with partial covering (2T-PC) yields $kT_{\rm low} = 14^{+6}_{-10}$~keV and $kT_{\rm high} = 43\pm12$~keV 
(lower left panel in Figure~\ref{fig:igr}) and we find that The 2T models are rather insensitive to the low-temperature value.
We note that the measured $kT_{\rm low}$ is consistent with the mean ionization temperature of several field IPs measured by \citet{Ezuka1999},  $kT_{\rm ion} = 8.1$~keV, as well as the $kT_{\rm low} 
\approx 5$~keV from our simulation in \S\ref{sec:primary_model}. 

\subsubsection{IP mass model fitting}

% IPM model fit 
While fitting an uncovered IPM model in the 3-50 keV band gives an unacceptably high $\chi_\nu^2 = 3.9$, it fits well to the 
15-50~keV \nustar\ data 
giving a WD mass measurement
of $M_{\rm WD} = 1.16\pm0.05 M_\odot$ with $\chi_\nu^2 = 1.02$. 
The IPM model with a partial covering (IPM-PC) gives an excellent $\chi^2_\nu=1.00$ and a tightly constrained WD mass of
$M_{\rm WD} = 0.98\pm0.03 M_\odot$ (lower right panel in Figure~\ref{fig:igr}).
The partial covering column density and fraction are in reasonable agreement with those seen in the 1T model.
The IPM model with reflection (IPM-RFL) gives a good $\chi_\nu^2 = 1.04$, ignoring the energy range 5.5-7.5~keV, but has a best-fit value
of ${\tt rel_{refl}}$ greater than unity.
With this value is frozen to one, the IPM model indicates an extreme WD mass of $M_{\rm WD} = 1.34\pm0.02M_\odot$. 

The unphysical value of {$\tt rel_{refl}$} obtained above motivates a hybrid model with both reflection and a partial 
covering, as shown in Table~\ref{tab:igr_fit}. 
This model gives sensible results for {$\tt rel_{refl}$}, observation angle, abundances and covering fraction when IPM was used as the 
source spectrum, and yields a lower WD mass of $M_{\rm WD} = 1.05^{+0.03}_{-0.02} M_{\odot}$. 
These masses are in much better agreement with those seen using only the partial covering component. 
The lower partial covering column density and covering fraction in the hybrid model demonstrate that the 
surface reflection has much the same effect as an accretion curtain. 
It is possible that further analysis of phase-resolved data will allow for better assessment of the relative contributions of intrinsic material and 
reflection to the emission spectrum. 
However for this phase-averaged data, using a partial covering model alone provides good fits to the spectrum.

%%%%%%%%%%%%%%%%%%%%%%%%%%%%%%%%%%%%%%%%%%%%%%%%%%%%%%%%%%%%%%%%%%%%%%%%%%%%

\section{Systematics associated with white dwarf mass and temperature measurements}
\label{sec:systematics}

The temperatures and masses for TV Col and J17303 found in our \nustar\ analysis are compared with those found by other missions in Table~\ref{tab:compare_kT_mass}.
When similar spectral models are used, the \nustar\ results are consistent with these previous observations.
A comparison of the various models, however, indicates that WD mass and temperature estimates have systematic uncertainties, largely due to the limited bandwidth (\S\ref{sec:limited_band}) and the effects of X-ray reprocessing components (\S\ref{sec:wd_mass}), 
that are generally larger than
the statistical uncertainties quoted by previous analyses. 

\begin{deluxetable*}{lccccccc}[]
\tablecaption{Individual IP comparison of temperature and white dwarf mass between instruments}
\tablecolumns{8}
\tablehead
{  \colhead{} & \colhead{}  & \colhead{} & \multicolumn{2}{c}{TV Col}   & \multicolumn{2}{c}{J17303} & \colhead{}  \\
\colhead{Instrument}  & \colhead{Model} & \colhead{Energy Band} & \colhead{Temperature} & \colhead{Mass} & \colhead{Temperature} & \colhead{Mass} & \colhead{Reference} \\
\colhead{ } & \colhead{ } & \colhead{keV} & \colhead{keV} & \colhead{$M_\odot$} & \colhead{keV} & \colhead{$M_\odot$} & \colhead{ } }
\startdata
\nustar\        & 1T      & 3-50     & $20.7\pm0.4$   & ---          & $64.0_{-0.5}$    & ---          & This work        \\
\asca\          & 1T  & 5-10     & $18^{+14}_{-6}$   & ---          & ---          & ---          & \citet{Ezuka1999} \\
\swift\         & 1T  & 14-195   & $21.6\pm2.4$  & ---    & $37.1\pm4.4$   & ---          &  \citet{Brunschweiger2009} \\
\integral\      & 1T  & 20-100   & ---  & ---                     & $31.6^{+12.7}_{-7.8}$   & ---          & \citet{Landi2009} \\
\nustar\        & 1T-PC/RFL\tablenotemark{a} & 3-50   & $16.1\pm0.5$   & ---          & $34\pm2$    & ---          & This work        \\
\nustar\        & 2T-PC\tablenotemark{b}   & 3-50     & $19.7^{+0.9}_{-0.8}$   & ---          & $43\pm12$  & ---          & This work        \\
\nustar\        & IPM     & 15-50    & ---  & $0.77\pm0.03$  & ---   & $1.16\pm0.05$   & This work \\
\nustar\        & IPM-PC/RFL\tablenotemark{a}  & 3-50 & ---  & $0.79\pm0.05$  & ---   & $1.1_{-0.1}^{+0.2}$   & This work        \\
\swift\         & IPM     & 14-195   & ---  & $0.78\pm0.06$  & ---   & $1.08\pm0.07$         &  \citet{Brunschweiger2009} \\
\asca\          & Fe line ratio\tablenotemark{c}  & 5-10 & ---          & $0.51_{-0.22}^{+0.41}$ & ---          & ---          & \citet{Ezuka1999} \\
\suzaku\        & IP-PSR-PC\tablenotemark{d}  & 2-40     & ---  & $0.91^{+0.14}_{-0.10}$  & ---   & $1.06^{+0.19}_{-0.14}$   & \citet{Yuasa2010} \\
\suzaku\        & IP-PSR\tablenotemark{d}  & 15-40    & ---  & $0.87^{+0.53}_{-0.18}$  & ---   & $1.20^{+0.15}_{-0.22}$   & \citet{Yuasa2013}
\enddata
\tablecomments{\nustar\ model descriptions can be found in Table~\ref{tab:models} and in Section~\ref{sec:spec_models}. The IPM model is based on the \citet{Suleimanov2005} mass model.}
\tablenotetext{a}{We fit the two spectral models (IPM-PC and IPM-RFL), and list the mean of their best-fit parameters. In addition to statistical errors,
systematic errors are adopted from the differences in the best-fit parameters between the two models.}
\tablenotetext{b}{The higher temperature values are listed here.}
\tablenotetext{c}{\citet{Ezuka1999} calculated ionization $kT$ from the measured intensity ratio of He-like and H-like Fe lines and determined $M_{\rm WD}$ using the \citet{Aizu1973} model.}
\tablenotetext{d}{IP-PSR model is a 1D accretion flow modelling the continuum and Fe lines, and is described in \citet{Yuasa2010}. IP-PSR-PC stands for IP-PSR model with partial covering absorption.}
\label{tab:compare_kT_mass}
\end{deluxetable*}

\subsection{The narrow 3-10 keV band spectral fitting and ionization temperature}
\label{sec:limited_band}

As described in \S\ref{sec:primary_model}, we use the narrow energy band of 3-10~keV and a partially covered 1T model to extract our effective ionization temperature. 
\nustar\ 3-10~keV data fit by a 1T-PC model produces a mean ionization temperature $kT_{\rm ion} = 9.8 \pm0.9$ (TV~Col) and $10.2^{+3.5}_{-1.1}$~keV (J17303). For comparison, 
using the Fe line ratios measured by \asca\ data, \citet{Ezuka1999} measured $kT_{\rm ion} = 7.7\pm2.7$~keV for TV~Col. 
However, the ionization temperature derived from soft X-ray band fitting or Fe line ratios is not a reliable WD mass estimate for several reasons. 

Since soft X-ray emission originates from near the bottom of the accretion column, 
the single ionization temperature 
does not represent the continuum (shock) temperature 
which is proportional to the WD mass. 
If the ionization temperature can be misinterpreted as the continuum temperature of the accretion column, 
the implied WD mass of the IP population is grossly underestimated, thus it 
requires an additional model-dependent step to derive WD mass.  

\citet{Ezuka1999} used the model relation developed by \citet{Aizu1973} to convert $T_{\rm ion}$ to 
$M_{\rm WD}$. However, this derivation can produce large errors associated with  $M_{\rm WD}$ due to the 
large $dM_{\rm WD}/dT_{\rm ion}$ gradient in the model for the relevant temperature range as well as 
the unknown temperature at the base of the accretion column ($kT_{\rm B}$). Using the Aizu model with $kT_{\rm B} = $ 0-2~keV, our ionization temperature 
for TV~Col implies a
WD mass of $M_{\rm WD} =  0.8\pm0.1 M_\odot$ which is consistent with that determined from the broad-band 3-50~keV continuum fits, 
whereas a lower mass of $M_{\rm WD} = 0.51^{+0.42}_{-0.22} M_\odot$ was derived from \citet{Ezuka1999}. 
The extreme sensitivity to $kT_{\rm ion}$ measurement is evident as only a small $kT_{\rm ion} \approx 2$~keV difference between the \nustar\ and \asca\ results  
leads to WD mass discrepancy as large as $\Delta M_{\rm WD} \approx 0.3 M_{\odot}$.  
This sensitivity issue similarly affects \nustar\ 3-10~keV spectral fitting for J17303 where $kT_{\rm ion} = 10.2^{+3.5}_{-1.1}$~keV leads to $M_{\rm WD} = 0.8_{-0.1}^{+0.6} M_\odot$, 
while the broad-band spectral fit with IPM models yield a higher WD mass $M_{\rm WD} = 1.16 \pm 0.12 M_{\odot}$ with significantly smaller errors.  

Overall, the mean WD mass measured in \citet{Ezuka1999} is systematically underestimated by $\sim0.3M_\odot$, compared to the broad-band fit. 
The mean ionization temperature of the 13 field IPs was measured by \citet{Ezuka1999} to be $kT \approx 8.8$~keV.
This corresponds to a mean WD mass of $0.55~M_\odot$, which is considerably less than the mean field IP mass of $0.86 \pm0.07M_\odot$ measured by \integral\ and \xmm\ 
\citep{Bernardini2012} and the mean mass of the 24 IPs measured by \citet{Yuasa2010}, $0.88\pm0.25M_\odot$. 
The narrow energy band of \asca, with its sharply decreasing effective area above 7~keV, makes it difficult to accurately estimate the soft X-ray continuum,
leading to a large uncertainty in the temperature and a best-fit WD mass that is much lower than other observation. 
%Thus we conclude that the soft X-ray band fit is an unreliable estimator of the WD masses, not only for the Fe line ratios from which \asca\ measured a lower WD mass, but also for the 3-10 keV \nustar\ spectroscopy (fitting both the continuum and Fe emission lines) with a higher, flatter effective compared to \asca, \chandra\ and \xmm. 

Thus we conclude that soft X-ray (narrow) band fits are an unreliable WD mass estimator.
The narrow band fits underestimate ionization temperature, leading to systematically lower WD masses,
or masses with very large errors.
This conclusion is clear for the \asca\ results on IP masses compared to \nustar\ and \xmm/\integral.
This conclusion will also pertain to narrow band temperature measurements with \chandra\ or \xmm.  But \xmm\ 
plus \integral\
results, broadband \suzaku\ results, and broad band \nustar\ results all produce good agreement on WD masses.

\subsection{White dwarf mass measurements}
\label{sec:wd_mass}

% Full-band spectral fitting with IPM-PC or IPM-RFL

It is evident that including a partial covering or reflection model systematically decreases the measured temperature and WD mass for TV Col and 
J17303. The temperature and WD mass are very sensitive to the partial covering column density, thus they can be largely overestimated if the 
X-ray reprocessing is significant but it is not taken into account for spectral fitting. 
Therefore, the most physically well motivated model, an IPM model with partial covering and/or reflection, 
that is fit over a broad energy band provides the most robust mass estimate. 
Although the IPM-PC and IPM-RFL models show nearly
identical spectral shapes thus yielding similar fit quality, they give small but finite differences in the fit WD mass values.
Then, we take a range of the best-fit WD masses by different spectral models as systematic errors. As shown in Table~\ref{tab:compare_kT_mass},
the systematic errors due to applying different models dominate over the statistical errors.

For TV~Col, the IPM model fits yield $M_{\rm WD} = 0.79\pm0.05 M_\odot$ for TV~Col where the systematic errors corresponding to
the difference in the best-fit $M_{\rm WD}$ values 
between the IPM-PC and IPM-RFL model are added to the statistical errors. 
For J17303, the 3-50~keV band fits with the three IPM models (IPM-PC, IPM-RFL with $\tt rel_{refl}$ fixed to 1, and IPM-PC-RFL model), that are statistically
acceptable and physically reasonable, give the mean WD mass of $M_{\rm WD} = 1.1_{-0.1}^{+0.2} M_\odot$. 
Given that our results are in good agreement with the \suzaku\ results \citep{Yuasa2013} where the Fe lines were fit between 6 and 7 keV, 
the \nustar\ observations establish that broad-band 3-50~keV observations are effective in constraining temperatures and masses without the need to
fit the 5.5-7.5~keV soft X-ray line emission region.

% E > 15 keV fitting 
On the other hand, fitting only the $E>15$~keV spectrum yields a WD mass, 
$M_{\rm WD} = 0.77\pm0.05$ (TV~Col) and $1.16\pm0.05 M_\odot$ (J17303), consistent with the broad-band fit to an IPM model with partial covering or reflection.
Although the limited bandpass induces larger statistical errors, 
this approach gives another robust measurement of WD mass as well as $kT_{\rm high}$ in the 2T model since it removes the dependence on partial covering and/or 
reflection which are neglibile at $E>15$~keV. 
This is further supported by the results of fitting our simulated partially covered IPM model with a range of IP mass $M_{\rm WD} = $0.8-1.0$M_\odot$ and column
density up to $N_H^{\rm pc}\sim 3\times10^{24}$~cm$^{-2}$ with an uncovered IPM model at high energies, which returns the correct WD mass.
Based on our simulation and \nustar\ data fitting, we find that the $E>15$~keV band for the two IPs is optimal since partial covering does
not affect the WD mass measurement and statistical errors remain small.

For each of TV~Col and J17303, we obtain the consistent WD masses between the two fitting methods, and they are in good agreement 
with the previous measurements (Table~\ref{tab:compare_kT_mass}). Similarly,
using a large sample of field IPs and their \suzaku\ data, \citet{Yuasa2013} demonstrated
that broad-band spectral fits  with a partially covered IP mass model and high energy
($E>15$~keV) fits with an uncovered IP mass model yield essentially the same WD masses within statistical uncertainties.
Our results for TV~Col and J17303 are consistent with the \integral\ and \swift\ analysis
extending to $E\ga100$~keV \citep{Landi2009, Brunschweiger2009}, thus our 3-50 keV band with \nustar\ is sufficiently ``broad'' to measure WD masses
accurately.
In summary, a combination of the two fitting methods with \nustar\ broad-band X-ray spectroscopy yields the most accurate measurement of WD masses.

%%%%%%%%%%%%%%%%%%%%%%%%%%%%%%%%%%%%%%%%%%%%%%%%%%%%%%%%%%%%%%%%%%%%%%%%%%%%%%

\section{Spectral analysis of the CHXE}
\label{sec:CHXE}

The results of our analysis of TV Col and J17303 lead us to conclude that narrow-band low-energy measurements can underestimate WD masses,
whereas high-energy measurements of the continuum temperature lead to robust mass estimates. 
Thus, a re-evaluation of the origin of the CHXE and its relation to the point source populations in the larger GC region is warranted, and we perform a 
reanalysis of \xmm\ and \nustar\ data from the CHXE using the spectral models discussed above.  

\subsection{Thermal plasma model fitting}

In order to compare our measurements of the CHXE with the soft X-ray measurements of the larger GC region,   
 we fit only 2-10~keV \xmm\ spectrum of the southwest CHXE region (as defined in KP15) with the same two-temperature model with no partial covering 
that was used in previous \chandra\ \citep{Muno2004} and \xmm\ (HW13) studies of the surrounding GC region.
In these previous analyses, the lower temperature accounts for a diffuse plasma with temperature $kT_1 \approx 1$~keV and the higher temperature of
$kT_2 \approx $7-8~keV is interpreted as arising from a population of magnetic CVs with mean WD mass  $M_{\rm WD} \approx 0.5 M_\odot$.
In our analysis, we recover a $kT_2 = 7.7^{+0.8}_{-1.0}$~keV that is consistent with these previous works. See Table~\ref{tab:compare_temperature} listing a comparison of 
the temperatures of the GC diffuse emission 
derived from the various models applied in different energy bands. 
The abundance for the high-temperature component, presumably associated with accretion onto the WD, is approximately solar, and the column density is
consistent with previous GC measurements.
%When applying a partial covering model to the CHXE, the derived temperature decreases to $kT_2 = 5.5^{+2.4}_{-0.9}$~keV, and the abundance decreases from nearly solar to a sub-solar value, more consistent with that of individual IPs \citep{Yuasa2010, Landi2009, Bernardini2012}.  
%The partial covering column density and fraction are poorly constrained, but the lower limit on the partial column density is $\sim3$ times that of the interstellar absorption. 
%This temperature is consistent with that seen in our simulations of IPs with $M_{\rm WD} = 0.8$-$1 M_\odot$ (see \S~\ref{sec:primary_model}). 
%Thus we conclude that using a partial covering model in the CHXE region, and by inference also in the HW13 region, drives down the intermediate temperature by several keV.  

\begin{deluxetable*}{lllllll}
\tablecaption{Galactic Center plasma temperature comparison between this work and previous measurements}
\tablewidth{0pt}
\tablecolumns{7}
\tablehead { \colhead{Region or Source} & \colhead{X-ray telescope}  & \colhead{Energy Band [keV]} &  \colhead{Models} & \colhead{$kT_2 $ [keV]} & 
\colhead{$kT_3$ [keV]} & \colhead{Reference} }
\startdata
SW region in CHXE     &  \xmm\ only     & 2-10         & 2T      & $7.7_{-1.0}^{+0.8}$   & ---              & This work \\
just outside CHXE     &  \xmm\          & 2-10         & 2T\tablenotemark{a}  & 7.5             & ---              & HW13      \\
SW region in CHXE     &  \nustar\       & 2-40         & 3T      & $7.2_{-1.3}^{+1.4}$  & $58_{-23}^{+127}$      & KP15     \\
%SW region in CHXE     &  \nustar\       & 2-40         & 3T-PC   & 4.8-5.8         & $43_{-26}$ & This work  \\
$2^\circ \times 0.8^\circ$ GC field & \chandra\ & 2-10 & 2T\tablenotemark{a} & $\ga 8$  & --- & \citet{Muno2004}
\enddata
\tablecomments {All models presented in this table fit two or three thermal components. The lowest temperature component fits to $kT_1\sim1$ keV diffuse emission, and it is 
not listed here. $kT_2$ refers the higher temperature of the 2T models or the middle temperature of the 3T models. $kT_3$ refers to the highest temperature of 
the 3T models.}
\tablenotetext{a}{HW13 fit thermal bremsstrahlung model, while the XSPEC model {\tt mekal} \citep{Mewe1986} was used by \citet{Muno2004}. In the both cases, the lower 
temperature component in their 2T models represent $kT \sim 1$ thermal diffuse emission. }
\label{tab:compare_temperature}
\end{deluxetable*}

Using broad 2-40~keV energy band data, KP15 fit the joint \xmm\ and \nustar\ spectrum of the southwest CHXE region to a three-temperature model with no partial covering,
yielding $kT_1 = 1.0^{+0.3}_{-0.4},                                                                                                                                 
kT_2 = 7.5^{+1.6}_{-1.3}$ and $kT_3 = 58^{+127}_{-23}$~keV. In this model, one temperature represents the diffuse, cool plasma, and the other two represent the standard two-temperature component 
used for TV~Col and J17303 above. The first ($kT_1 = 1.0$~keV) and second temperatures ($kT_2 = 7.5$~keV) agree with the \xmm\ only measurement as well as 
\citet{Muno2004} and HW13. The third bremsstrahlung component, largely fitting the hard X-ray continuum, sets a lower limit of $kT > 35$~keV (KP15).  
%This model used two APECs to represent the cool $\sim1$~keV plasma and the line-emitting accretion region, and a bremsstrahlung component to represent the hard X-ray continuum.

\begin{deluxetable*}{lccccccc}
\tablecaption{Joint \xmm\ and \nustar\ spectral fitting results for the CHXE southwest region}
\tablewidth{0pt}
\tablecolumns{3}
\tablehead{\colhead{Parameters}  & \colhead{1T-IPM} & \colhead{IPM} }
\startdata
Instrument & XMM+\nustar\ & \nustar\  \\
Bandpass [keV] & 2-40 & 15-40 \\
$N^{f}_{\rm H} [10^{22}$\,cm$^{-2}$] & $14\pm1$ & --- \\
$kT_1$ [keV] & $1.0\pm0.2$  & --- \\
$Z_1$ & $5.0_{-3.2}$ & --- \\
$N_1$ [$10^{-2}$cm$^{-3}$] & $8.43^{+0.21}_{-0.02}$ & ---  \\
$M_{\rm WD}$ [$M_\odot$] & $0.87^{+0.11}_{-0.09}$ & $0.89^{+0.15}_{-0.12}$ \\
$N_{IPM}$ [$10^{-13}$] & $4.9^{+1.7}_{-1.3}$ & $4.8^{+2.5}_{-1.7}$ \\
Flux (2-10 keV)\tablenotemark{a} & $1.2$ & $1.2$  \\
Flux (20-40 keV)\tablenotemark{a} & $0.7$ & $0.7$ \\
$\chi_\nu^2$ (dof) & 1.07 (400) & 1.12 (52) 
\enddata
\tablecomments {$kT_1$ corresponds to the 1~keV thermal thermal diffuse emission in the GC and is unrelated to the IPs.}
\tablenotetext{a}{The flux unit is $10^{-12}$ \eflux.}
\label{tab:chxe_fit}
\end{deluxetable*}

\subsection{IP mass model fitting}

\begin{figure*}[t]
\mbox{
\epsfig{figure=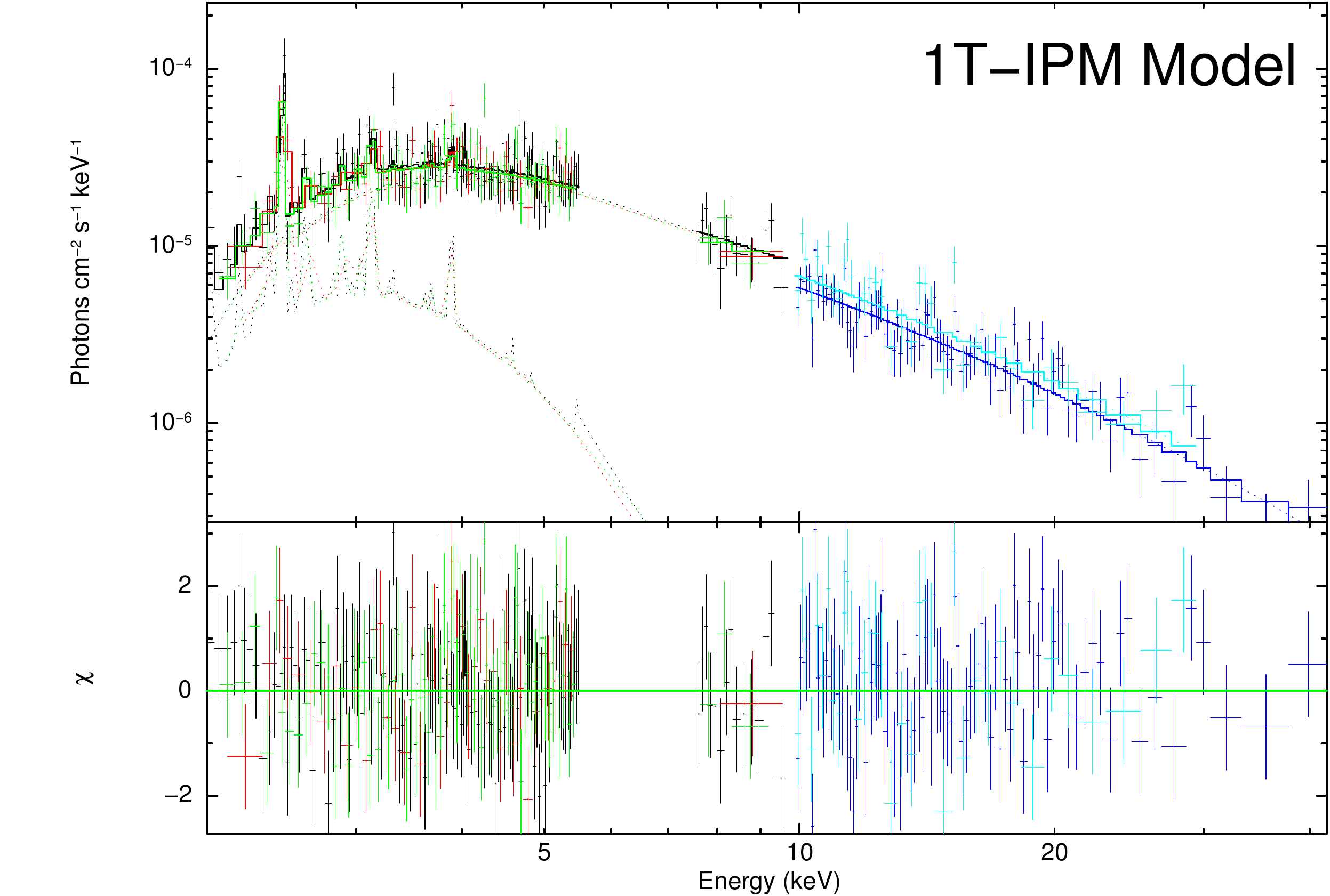,height=0.5\linewidth, angle=270}
\epsfig{figure=plots/chxeSW_above15keV_unfold.eps,height=0.5\linewidth, angle=270}
}
\caption{\xmm\ + \nustar\ 2-40~keV unfolded spectra of the CHXE southwest region (\xmm\ below 10~keV, \nustar\ above 10~keV) fit with 1T-IPM (lower panel, without 5.5-7.5 keV band) and IPM model (right panel, above 15 keV). }
\label{fig:chxe}
\end{figure*}

Following the analysis for TV~Col and J17303, we fit two additional spectral models to the CHXE data (Figure~\ref{fig:chxe} and Table~\ref{tab:chxe_fit}).
The 2-40~keV joint \xmm\ and \nustar\ spectra of the southwest CHXE region fit to a $kT_1 \sim 1$~keV APEC plus IPM model (1T-IPM), 
with the 5.5-7.5~keV region excluded and no partial covering, 
results in $\chi_\nu^2 = 1.07$ and mean WD mass of $M_{\rm WD} = 0.87^{+0.11}_{-0.09}M_\odot$.
The abundance of the cool plasma is poorly constrained, but consistent with previous measurements of the GC region.  
We also fit the $\sim1$~keV APEC plus IPM model with partial covering (1T-IPM-PC), as we argue the most physical model is the IPM model with
partial covering for the case of TV~Col and J17303.  
While the fit was acceptable with $\chi_\nu^2 \approx 1$, the partial covering column density for the CHXE region is poorly constrained, due to spectral contamination from the
diffuse $kT \approx 1$~keV plasma and the interstellar absorption in the region.

To check whether the model without partial covering is accurately measuring the mean WD mass, we performed a fit using \nustar\ 15-40~keV data.
For both TV Col and J17303, fits to the $ E >15$~keV X-ray continuum yield reliable masses and hard X-ray continuum temperatures that are consistent with both \nustar\ broad-band fits and measurements by other observatories (see Tables~\ref{tab:compare_kT_mass}, \ref{tab:tvcol_fit}, and \ref{tab:igr_fit}). 
At these high energies, a partial covering does not affect the measured temperature. 
The 15-40~keV \nustar\ fit to the CHXE southwest region yields a mean WD mass of $M_{\rm WD} = 0.89^{+0.15}_{-0.12}M_\odot$ (see Table~\ref{tab:chxe_fit}), 
in agreement with the 2-40~keV joint \xmm\ and \nustar\ fit with the 1T-IPM model. 
Similarly, we fit an uncovered IPM model to the 15-40~keV \nustar\ data of the northeast region of the CHXE (as defined in KP15). 
The best-fit WD mass from the northeast region is $M_{\rm WD} = 0.95_{-0.12}^{+0.14}M_\odot$, and 
it is consistent with that of the southwest region. However, the northeast region is contaminated by $kT\sim5$~keV thermal component from Sgr A East and non-thermal emission from several prominent X-ray filaments. Based on these considerations, we conclude that the most reliable mass estimate for the CHXE is obtained by 
fitting the southwest spectrum above 15~keV to an IPM model without partial covering, yielding a mass of  $M_{\rm WD} = 0.89^{+0.15}_{-0.12}M_\odot$. 
%This is consistent with the mass obtained by fitting an uncovered 1T-IPM model, which resulted in $M_{\rm WD} = 0.87^{+0.11}_{-0.10} M_\odot$.

%%%%%%%%%%%%%%%%%%%%%%%%%%%%%%%%%%%%%%%%%%%%%%%%%%%%%%%%%%%%%%%%%%%%%%%%%%
\section{Discussion}
\label{sec:discussion}

%\chandra\ and \xmm\ analysis indicated that the $\sim$50~pc region surrounding Sgr A* is also dominated by IPs, but with a mean mass of 
%$M_{\rm WD} \approx 0.5M_\odot$~\citep{MunoDiffuse2004, Muno2004, Heard2013A}. 
%This result is consistent with the \asca\ field IP measurements~\citep{Ezuka1999}, and was further supported by observations of the Galactic bulge and ridge, 
%which indicated populations with mean 
%$M_{\rm WD} \approx 0.48$-$0.66M_\odot$~\citep{Yuasa2012} and $M_{\rm WD} \approx 0.5M_\odot$~\citep{Krivonos2007}, respectively.
%However, the \chandra\ results should be viewed with caution due to the poor statistical quality of their identified  point sources. 
%Most importantly, the 2-10~keV spectral fit is extremely sensitive to the X-ray continuum at $>5$~keV, where the \chandra\ effective area is lowest, and is not able to 
%distinguish possible additional higher-temperature components. 

\subsection{The origin of the CHXE} 

In light of our analysis of TV Col, J17303, and the CHXE using a wide variety of IP spectral models, we propose that the CHXE is an unresolved population of IPs with 
mean WD mass $M_{\rm WD} \approx 0.9M_\odot$. We also list other observational evidences supporting the IP scenario.  
%Using broad energy-band data, the CHXE is consistent with one population of IPs with $M_{\rm WD} \approx 0.8$-$1.0M_\odot$, regardless of the spectral model that is applied. 

\begin{itemize} 

\item In the soft X-ray band, the southwest region of the CHXE outside the SNR Sgr A East shows $kT\approx8$~keV thermal emission with both neutral and ionized Fe lines in addition to the diffuse $kT\sim1$~keV component, indicating a 
dominant population of CVs. 
While milli-second pulsars, BH-LMXBs or non-thermal 
diffuse emission can account for the CHXE as a power-law spectrum with $\Gamma\sim$1.5-2 above 20~keV, their low-energy extensions below 10~keV are 
inconsistent with the observed thermal emission with Fe lines. 

\item The CHXE exhibits an elliptical profile similar to the surface brightness distribution of the nuclear star cluster (NSC) where a  
high stellar density is observed in the IR band with a total luminosity of 
$\sim10^7L_\odot$ and mass of $\sim10^7M_\odot$. 
Contrary to the previous IR studies, after correcting the effects of polycyclic aromatic hydrocarbon (PAH)
emission and extinction carefully, \citet{Schodel2014} revealed a NSC profile flattened along the Galactic plane and centered at Sgr A* with a half light radius of $4.2\pm0.4$~pc. The similarity between the two spatial distributions  
strongly suggests a stellar origin of the CHXE. 

\item The CHXE has spectral characteristics closely matching those of the field IPs observed by both \integral/\xmm\ \citep{Bernardini2012} and \suzaku\
\citep{Yuasa2010}, including the derived masses. 
On the other hand, both non-magnetic CVs and polars have too soft X-ray spectra to account for the 20-40 keV emission of the CHXE \citep{Cropper1998, Byckling2010}. The \integral/IBIS Galactic Plane survey detected one non-magnetic CV (SS~Cyg) and three polars, 
while 60\% of the known IPs were detected \citep{Barlow2006}.

\item The mean WD mass of the CHXE ($M_{\rm WD} \approx 0.9M_\odot$) is 
remarkably consistent with that of the field IPs \citep[$M_{\rm WD} =$0.86-0.88$M_\odot$,][]{Bernardini2012, Yuasa2010}.  
Also, the mean WD mass among all CVs measured by SDSS is $M_{\rm WD} = 0.83\pm0.24 M_\odot$ \citep{Zorotovic2011} in excellent
agreement with the $M_{\rm WD} \approx 0.9 M_\odot$ derived for the CHXE (Table~\ref{tab:compare_mass}). 

\end{itemize} 

We thus conclude that the CHXE is consistent with one population of IPs with $M_{\rm WD} \approx 0.9 M_\odot$. 
The dominant $kT \approx 8$~keV component detected in the soft X-ray band corresponds to the lower temperature ($kT_{\rm low}$) present in                  
two-temperature model fits to 1-dimensional accretion flow models of IPs with $M_{\rm WD} \approx 
0.9 M_\odot$, where it is associated with the               
emission near the base of the accretion column or with the ionization temperature in the line emission region.  
                    
The lower mass $M_{\rm WD} \approx 0.5M_\odot$ obtained by \chandra\ and \xmm\ data \citep{MunoDiffuse2004, Muno2004, Heard2013A} can be explained as an artifact of limited, low energy-band spectral fitting.
As we demonstrated in \S\ref{sec:limited_band}, the narrow-band soft X-ray measurements underestimate the IP masses, and a precise measurement of the mean WD mass from the CHXE region
(which is unresolved from the larger GC region by \integral\ and \suzaku) is only possible through \nustar.

Recently it has been suggested that highly magnetized CVs are formed in a common envelope phase \citep{Tout2008}. This has been posited to explain why high magnetic field white dwarfs
(HMFWD) are more massive on average than the general WD population. The mean mass of the HMWFD, $M_{\rm WD} \approx 0.8 M_\odot$
\citep{Kawka2007, Ferrario2015}, is strikingly similar to that measured for the
IP population comprising the CHXE. Since the IPs are believed to represent an alternate endpoint of the common envelope phase, it may be natural that they are more
massive as well. This makes them hotter and thus easier to detect in hard X-ray observations.

\begin{deluxetable*}{llllll}
\tablecaption{Mean white dwarf mass comparison between this work and previous measurements}
\tablewidth{0pt}
\tablecolumns{6}
\tablehead { \colhead{Region or Source}  &  \colhead{X-ray Telescope} & \colhead{Energy Band [keV]} & \colhead{Models\tablenotemark{a}} & \colhead{$M_{\rm WD}$ [$M_\odot$]} & \colhead{Reference} }
\startdata
SW region in CHXE     & \nustar\             & 2-40        & 1T-IPM                    & $0.87_{-0.09}^{+0.11}$      & This work \\
SW region in CHXE     & \nustar\             & 15-40       & IPM                       & $0.89_{-0.12}^{+0.15}$     & This work  \\
NE region in CHXE     & \nustar\             & 15-40       & IPM                       & $0.95_{-0.12}^{+0.14}$    & This work \\ 
100~pc $\times$ 100~pc region around \sgra\ & \xmm\ & 2-10    & IP-PSR\tablenotemark{b}   & $0.49\pm0.02$    & HW13 \\
Galactic ridge        & \suzaku\             & 15-50        & IP-PSR\tablenotemark{b}   & $0.66_{-0.07}^{+0.09}$     & \citet{Yuasa2012} \\
Galactic ridge        & \integral\           & 20-200       & IPM                       & $0.60\pm0.05$          & \citet{Turler2010} \\
Field CVs                 &       ---               &  ---           &      ---                &  $0.83\pm0.24$ &  \citet{Zorotovic2011} \\ 
Field IPs             & \integral\ \& {\it XMM}   & 15-100     & IPM                      & $0.86\pm0.07$     & \citet{Bernardini2012} \\
Field IPs             & \suzaku\             & 2-40        & IP-PSR\tablenotemark{b}   & $0.88\pm0.25$     & \citet{Yuasa2010} \\ 
Isolated DA WDs          &       ---               &  ---           &      ---                &    $0.593\pm0.016$             &  \citet{Kepler2007}   \\
Isolated high magnetic field WDs          &     ---               &  ---           &      ---                &    $0.784\pm0.047$         &  \citet{Ferrario2015}   
%\tvcol\               & \nustar\             & 3-50        & IPM-PC                    & 0.80 (0.78-0.81)     & This work. \\
%\tvcol\               & \nustar\             & 15-50       & IPM                       & 0.77 (0.74-0.80)     & This work. \\
%\igrsrc\              & \nustar\             & 3-50        & IPM-PC                    & 0.98 (0.95-1.01)     & This work. \\
%\igrsrc\              & \nustar\             & 15-50       & IPM                       & 1.16 (1.11-1.21)     & This work.
\enddata
\tablecomments {\nustar\ model descriptions can be found in Table~\ref{tab:models}.}
\tablenotetext{a}{5.5-7.5 keV data bins are ignored in \nustar\ IPM model fittings.}
\tablenotetext{b}{IP-PSR model is a 1D accretion flow modelling the continuum and Fe lines, and is described in \citet{Yuasa2010}.}
\label{tab:compare_mass}
\end{deluxetable*}

%%%%%%%%%%%%%%%%%%%%%%%%%%%%%%%%%%%
\subsection{Luminosity function and density of IPs in the Galactic Center}
\label{sec:CHXE_interpretation}

We investigate 
the luminosity function of the IPs in the CHXE. 
We use the total 2-8 keV luminosity $L_X = 3.6\times10^{34}$~erg\,s$^{-1}$ in the CHXE excluding the $kT\sim1$~keV diffuse component (KP15), a power-law index for the luminosity distribution ($N(>L_X) = k L_X^{-\alpha}$) with 
$\alpha\sim$1-1.5 \citep{Muno2009, Yuasa2012} and an $L_X^{\rm max} \sim 3\times10^{33}$~erg\,s$^{-1}$ for the maximum IP luminosity \citep{Pretorius2014}. 
The results are quite insensitive to the precise $L_X^{\rm max}$, provided $L_X^{\rm min}/L_X^{\rm max} \ll 1$.  An upper bound on $L_X^{\rm min}$ is obtained from the observation 
that the CHXE point sources are completely unresolved by \nustar.  Given the \nustar\ angular resolution of $\sim1$~pc at the GC, this requires $\ga$ few hundred sources.  
This requires an XLF extending down to $L_X \la 5\times10^{31}$~erg\,s$^{-1}$. 
The corresponding log$N$-log$L$ would suggest that the CHXE is composed primarily of the \chandra\ point sources in the \nustar\ 
field of view. 
An estimation of the IP density in the GC also suggests that the XLF of the CHXE IPs has $L_{\rm min} \la 5\times10^{31}$~erg\,s$^{-1}$.  

We can estimate the IP density directly from the density of main sequence stars in the GC.
From~\citet{Pretorius2013}, we find that the central 10~pc contains $\sim 7500$ main sequence stars per pc$^{3}$.
The IP density is $n({\rm IP}) = 7500*f_{\rm ab}*f_{\rm B}*f_{\rm bin}*f_{\rm cv}*f_{\rm ip}$ pc$^{-3}$.
Here $f_{\rm ab} \sim 0.28$ is the fraction of main sequence stars with mass $\sim$4.5-10 M$_\odot$ (A and B stars) sufficient to form $M_{\rm WD} > 0.8M_\odot$
\citep{Ferrario2005}.
The fraction of A and B stars that will form highly magnetized WDs is estimated as $f_{\rm B} \sim 0.4$~\citep{Ferrario2005}.
A number $\sim$ 2 times smaller still can be consistent with the estimated number of isolated, highly magnetized WDs presented in
\citet{Ferrario2005}.
\citet{Hopman2009} estimated the binary fraction $f_{\rm bin}$ as a function of normalized radius from the GC.
Inside the influence radius, which is the radius within which the supermassive black hole dominates dynamical processes,
the binary fraction sharply drops. Outside the influence radius, \citet{Hopman2009} quotes a binary fraction $f_{\rm bin}\sim0.05$.
The influence radius is of the order of a few pc, so the fraction of the total mass at $r < 10$~pc inside the influence radius is small,
and so we estimate the binary fraction as $f_{\rm bin} \sim 0.05$.
We very crudely set the fraction of WD binaries that are CVs ($f_{\rm cv}$) to $\sim1$, but this is a crude upper limit,
since only WD in tight binaries will form CVs.
 The fraction of magnetic CVs among all CVs is $f_{\rm mcv} \sim 0.2$ \citep{Pretorius2013}.
Based on a sample of 30 magnetic CVs detected by the {\it ROSAT} Bright Survey, \citet{Pretorius2013} estimated $f_{\rm ip} \sim 0.4$ as the fraction of
IPs among the magnetic CVs in the solar neighborhood. This yields $n_{\rm ip} \approx $~2-4~pc$^{-3}$ where the lower
density assumes $f_{\rm B} \sim 0.2$.

These IP densities are in good agreement with the estimate of HW13, $n_{\rm ip} \sim$3-10~pc$^{-3}$, near the southwest region of the CHXE. KP15 estimated a higher IP density, $n_{\rm ip} \sim 15$~pc$^{-3}$, because 
there it was assumed that the XLF of IPs extended  well below $10^{31}$ erg\,s$^{-1}$. 
But an XLF of the CHXE extending to $\la 5\times10^{31}$~erg\,s$^{-1}$ gives an IP density of $\ga$ few pc$^{-3}$, consistent with a \nustar\ unresolved CHXE point source 
population, the rough first principles estimate above and HW13.  
It is tempting to regard these estimates as
conservative since accretion rate calculations \citep{Yuasa2012} would suggest that WD mass could have grown $\sim0.05$-$0.1 M_\odot$ over several Gyr.
Using an analytic initial (progenitor)-final (WD) mass function \citep{Catalan2008}, IPs with $M_{\rm WD} \sim 0.8$-$0.9 M_\odot$ would be
$\sim0.05$-$0.1 M_\odot$ lighter at formation, implying progenitor masses $\sim0.8$-$1 M_\odot$ lighter.
Thus a larger fraction of main sequence stars would be available to produce massive IPs. However the question of whether CVs actually gain
mass by accretion or suffer mass loss through nova cycles is complex and unresolved \citep{Wijnen2015}.

Previous \chandra\ observations suggested that a large population of the X-ray point sources in the GC region are magnetic 
CVs, and most likely IPs. 
\citet{Muno2004} classified a majority of thousands of X-ray point sources in the central $2^\circ\times0.8^\circ$ region as possessing hard power-law photon 
indices $\Gamma\la1$ indicating that they are mostly IPs. \citet{Hong2012} found that a low extinction region called the Limiting Window, 
at $\sim1.4^\circ$ south of the GC, contains a large population of magnetic CVs. 
 
Recently, an extensive \nustar\ survey detected 70 hard X-ray point sources over a 0.6~deg$^2$ region in the GC region, and nearly all of them have \chandra\ counterparts  
\citep{Hong2016}. Their 10-40~keV log$N$-log$S$ distribution indicates that a majority of the \nustar\ hard X-ray sources 
should have $kT \sim 20-40$~keV to match with the \chandra\ source distribution. More than a dozen bright  
\nustar\ sources exhibit broad-band X-ray spectra that resemble those of heavy IPs ($\sim 0.8-1 M_\odot$) 
with Fe lines and $kT \sim 20-40$~keV when fit with a 1T APEC model. 
The \nustar\ point sources detected above 10~keV represents the bright tail ($L_X \ga 4\times10^{32}$~erg\,s$^{-1}$ 
in the 3-10 keV band) of the \chandra\ source luminosity distribution or only $\sim1$\% of the GC \chandra\ source population in the GC. 

We can estimate the contribution of IPs to the point sources in the $2\times0.8$~deg$^2$ region under the assumption that the CHXE XLF applies. 
The XLF is normalized by the $\ga$35 sources (all with \chandra\ counterparts) that \citet{Hong2016} observed and claimed to be IPs, and 
by the ratio of areas surveyed (0.6~deg$^2$ and 1.6~deg$^2$ respectively).  
If the XLFs of the CHXE and the \chandra\ sources are comparable, $L_X^{\rm min} \la 5\times10^{31}$~erg\,s$^{-1}$ implies that 
$\ga20$-40\% of the \chandra\ sources are IPs. The range of our estimates is due to the uncertainty in the slope of log$N$-log$L$ curves. 
If the \nustar\ sources observed by \citet{Hong2016} are primarily IPs and/or if the XLF of the CHXE IPs extends down by another factor 
of $\sim2$ (which is close to the sensitivity limit of the \chandra\ GC survey), 
an overwhelming majority of the \chandra\ point sources in the GC could be IPs. 

%%%%%%%%%%%%%%%%%%%%%%%%%%%%%%%%%%%
\subsection{IPs and the Galactic ridge and bulge populations}
\label{sec:ridge_bulge}

Observations of the Galactic ridge and bulge have also suggested a dominant population of IPs.  
\citet{Yuasa2012}, using \suzaku\ data, fit the $E>15$~keV spectrum of the bulge with a 1-dimensional accretion flow model.
The derived mean mass was $M_{\rm WD} = 0.66^{+0.09}_{-0.07} M_\odot$, which is inconsistent with the mean mass we estimate for the CHXE and 
field IPs (Table~\ref{tab:compare_mass}). 
Similar mean WD mass ($M_{\rm WD} = 0.60\pm0.05 M_\odot$) was obtained by \citet{Turler2010} using the \integral\ Galactic ridge data. 
The \integral\ and \suzaku\ IP model fits yield consistent results with masses $\sim0.6$-$0.66 M_\odot$ even when the
energy band is restricted to $E \ga$15-20~keV, and as we demonstrated, such hard X-ray band fits should yield the most reliable WD mass measurements.
Note that using broad energy band fits (3-50~keV) to the \suzaku\ ridge observations, the mean observed mass decreased to $\sim0.48 M_\odot$, similar to 
the $0.5 M_\odot$ obtained from \integral\ observations of the Galactic ridge \citep{Krivonos2007}.

The lower mean IP mass measured for the Galactic ridge, compared to the GC, is puzzling.
As mentioned above, the higher IP mass in the GC is more consistent with observations of CV masses in general, and the HFMWD,
which are believed to share a common evolutionary origin with magnetic CVs.
Given that both magnetic and non-magnetic CVs have the same high mass $\sim0.8$-$0.9 M_\odot$, why would the IPs of the Galactic ridge have such
markedly lower masses? 
One clue could be the systematically lower temperatures or softer power-law spectra observed in the Galactic ridge.
This is also seen in the derived temperatures, with the CHXE being substantially hotter than the $kT \approx 15$~keV measured by \suzaku\ and \integral. 
Additionally, broad-band {\it RXTE} \citep{Valinia1998} and \suzaku\ \citep{Yuasa2012} fits using a power-law model demonstrated softer photon 
indices ($\Gamma = 2.3$ and $\Gamma = 2.8$, respectively) than 
corresponding fits to the CHXE ($\Gamma = $1.2-1.9, KP15). 

What is required to account for the softer ridge emission is a dominant contribution from non-magnetic CVs and polars since their spectra 
are softer on average than IPs \citep{Eracleous1991, Cropper1998, Byckling2010, Reis2013}.  If the IPs went from being the dominant 
CV component in the GC to a sub-dominant component, compared to non-magnetic CVs and polars, in the Galactic ridge, then a spectral softening 
would result. 
Applying the IP model, as \suzaku\ and \integral\ did, to a spectrum dominated by softer non-magnetic CVs and polars, would lead to a systematically lower WD mass. 
Indeed, fitting a single optically-thin thermal plasma model to the X-ray spectra of local non-magnetic 
CVs yields $kT \la 10$~keV, which is significantly softer than TV~Col ($kT \approx 20$~keV) and most IPs \citep{Byckling2010}. 

Recently \citet{Reis2013} found an average X-ray luminosity of 20 optically selected non-magnetic CVs is 
$8\times10^{29}$~erg\,s$^{-1}$, contrary to the previous studies whose samples were biased toward selecting CVs 
with higher accretion rates. It indicates that a large population of non-magnetic CVs may remain unresolved at a distance of 
8~kpc but they may have significant contribution to the Galactic ridge diffuse X-ray emission. Indeed, based on detailed Fe line 
diagnostics, \citet{Xu2016} conclude that the Galactic ridge X-ray emission may consist largely of non-magnetic CVs, not polars, while IPs 
contribute to a population of brighter and harder sources.  
Detailed synthetic modeling of the various source populations, taking into account the differences between the ridge 
and GC regions, such as metallicity, could also prove useful.

%%%%%%%%%%%%%%%%%%%%%%%%%%%%%%%%%%%%%%%%%%%%%%%%%%%%%%%%%%%%%%%%%%%%%%%%%%%%%%%%%%%%%%%%%%
%%%%

\section{Summary}
\label{sec:summary}

\begin{itemize} 

\item[(1)] {\it NuSTAR} observations of TV~Col and J17303 have demonstrated that the most robust masses are obtained using a model that 
incorporates 1-dimensional accretion flow approximations and absorption through the accretion curtain. 
Restricting the energy band of the model fits to $E \ga 15$~keV and ignoring absorption produces equally reliable masses. 

\item[(2)] The \nustar\ observations reveal that the emission from the CHXE and bright \chandra\ X-ray point sources 
($L_X \ga 4\times10^{32}$~erg\,s$^{-1}$) in the Galactic Center is consistent with a large population of 
IPs with mean mass of $M_{\rm WD} \sim0.9M_\odot$.  
This is a very natural number, since it is in good agreement with the mean mass measured 
for all CVs by SDSS \citep{Zorotovic2011}. 

\item[(3)] The softer diffuse X-ray emission with $kT \approx $7-8~keV detected by \chandra\ and \xmm\ and the harder CHXE have a common origin in massive (compared to isolated WD) IPs.

\item[(4)] 
The CHXE XLF must extend down to a luminosity of $L_X \la 5\times10^{31}$~erg\,s$^{-1}$ in order for the IPs to remain unresolved by 
\nustar. The density of IPs derived from this XLF is consistent with previous \xmm\ observations, and with crude estimates based on the mass density for the GC, initial-final mass 
function for WD, and probability that these WDs end up in IPs. 

\item[(5)] 
If the XLFs of the CHXE and the \chandra\ sources are comparable, $L_X^{\rm min} \la 5\times10^{31}$~erg\,s$^{-1}$ implies that
$\ga20$-40\% of the \chandra\ sources are IPs. 
If the \nustar\ sources observed by \citet{Hong2016} are primarily IPs and/or if the XLF of the CHXE IPs extends down by another factor
of $\sim2$, an overwhelming majority of the \chandra\ point sources in the GC could be IPs.

\item[(6)] The Galactic ridge X-ray emission is much softer than the CHXE, and thus may not be composed primarily of IPs. 
It likely contains an admixture of non-magnetic CVs and polars, with the harder tail comprised of IPs. 
This is similar to the conclusion of a recent analysis of Fe-line emission from the ridge \citep{Xu2016}, 
who proposed that DNe are the bulk of the soft X-ray emission and IPs the origin of the harder emission.

\end{itemize} 

\acknowledgements

This work was supported under NASA Contract No. NNG08FD60C, and made use of data from the \nustar\ mission, a project led by the California Institute of 
Technology, managed by the Jet Propulsion Laboratory, and funded by the National Aeronautics and Space Administration. We thank the \nustar\ Operations, 
Software and Calibration teams for support with the execution and analysis of these observations. This research has made use of the \nustar\ Data Analysis 
Software (NuSTARDAS) jointly developed by the ASI Science Data Center (ASDC, Italy) and the California Institute of Technology (USA). The authors thank 
K. Mukai and Q.D. Wang for valuable discussions. 

\bibliography{NuSTAR_IP}

\appendix 

\begin{figure*}[t]
%\centerline{                                                                                                                                                                                                                                               
\mbox{
\epsfig{figure=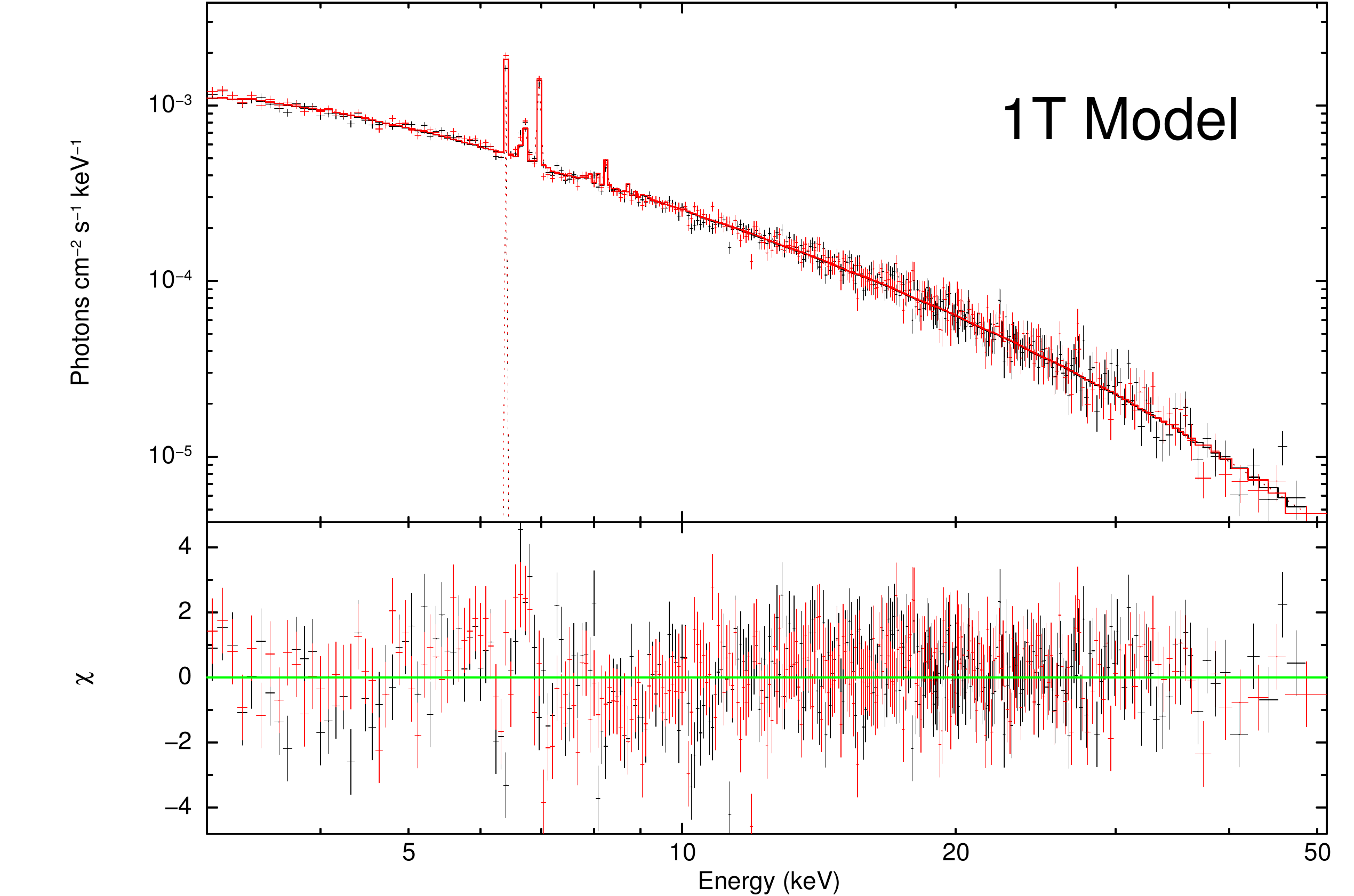,height=0.5\linewidth, angle=270}
\epsfig{figure=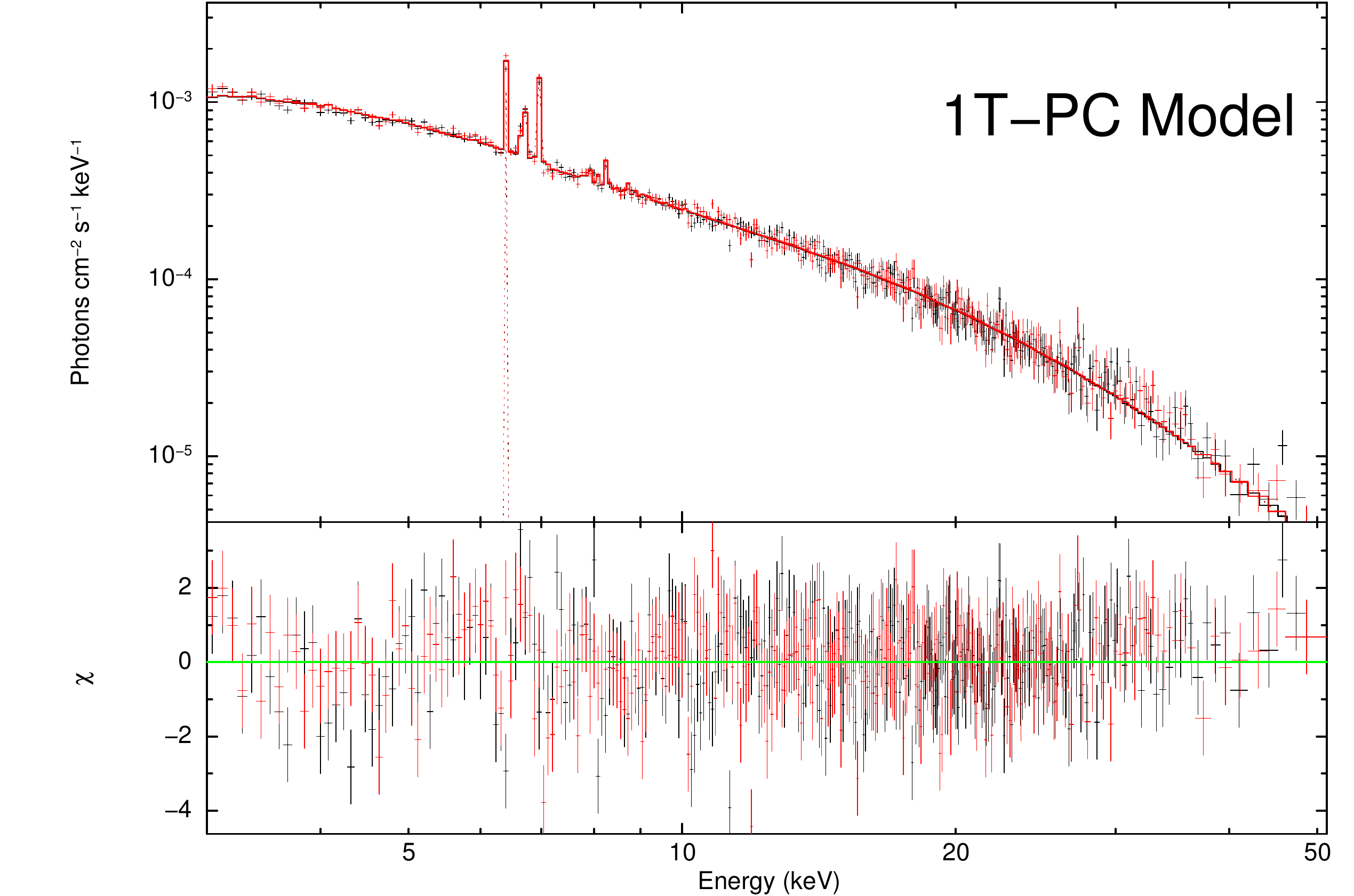,height=0.5\linewidth, angle=270,}
}
\mbox{
\epsfig{figure=plots/tvcol_2t_kT1free_unfold.eps,height=0.5\linewidth, angle=270}
\epsfig{figure=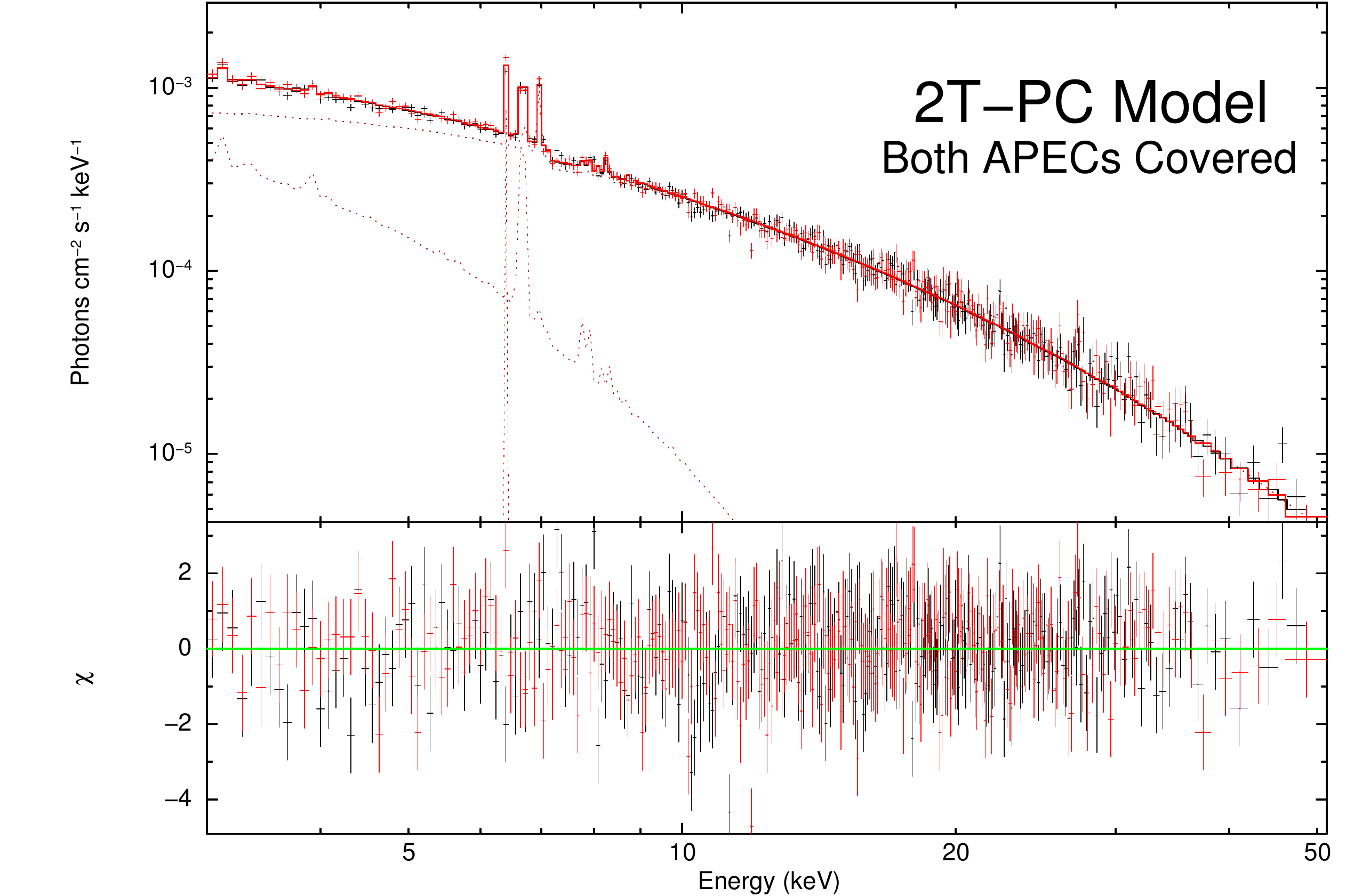,height=0.5\linewidth, angle=270}
}
\mbox{
\epsfig{figure=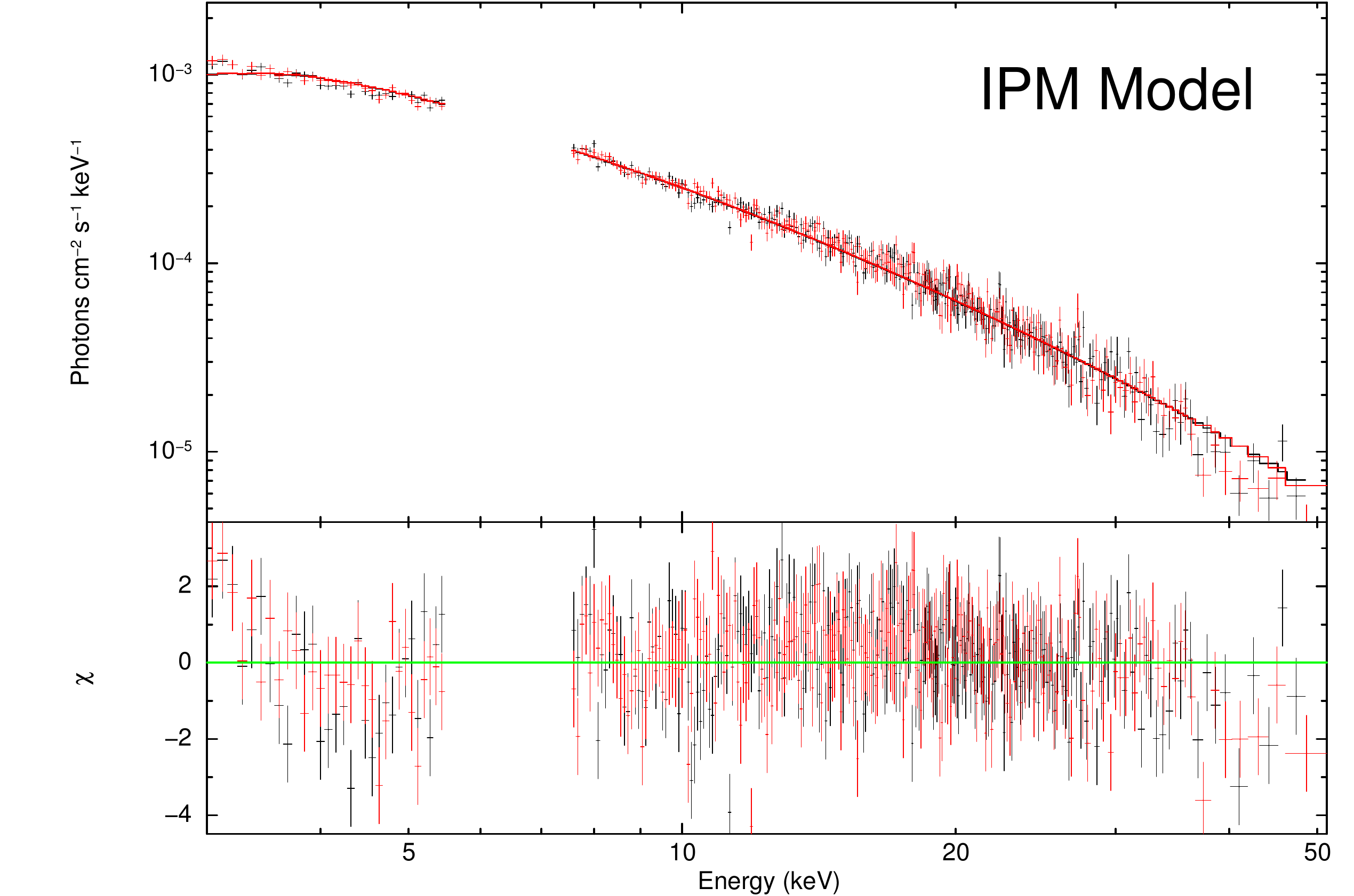,height=0.5\linewidth, angle=270}
\epsfig{figure=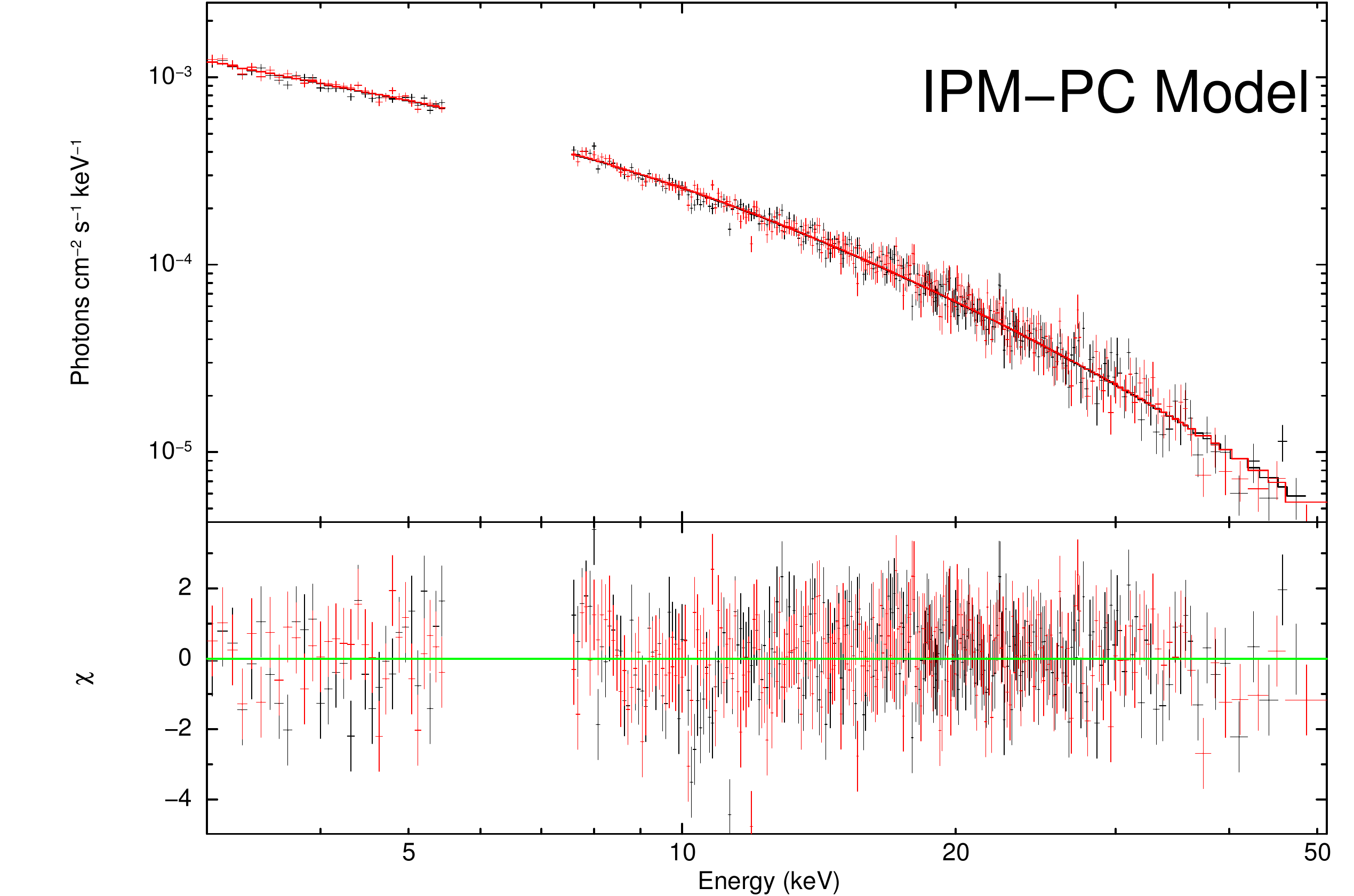,height=0.5\linewidth, angle=270}

}
\caption{\nustar\ 3-50~keV unfolded spectra of \tvcol\ (Black: FPMA, Red: FPMB) fit with 1T models (top panels), 2T models (middle panels)
and IPM models by ignoring the Fe line complex in the 5.5-7.5 keV band (bottom panels). The left and right panels show the spectral fits without and with partial-covering absorption, respectively.
The spectral fits with the reflection model are not presented here since they show similar residuals and fit quality as the partially covered models (Table~\ref{tab:tvcol_fit}).}
\label{fig:tvcol}
\end{figure*}

\begin{figure*}[t]
%\centerline{                                                                                                                                                                                                                                              
\mbox{
\epsfig{figure=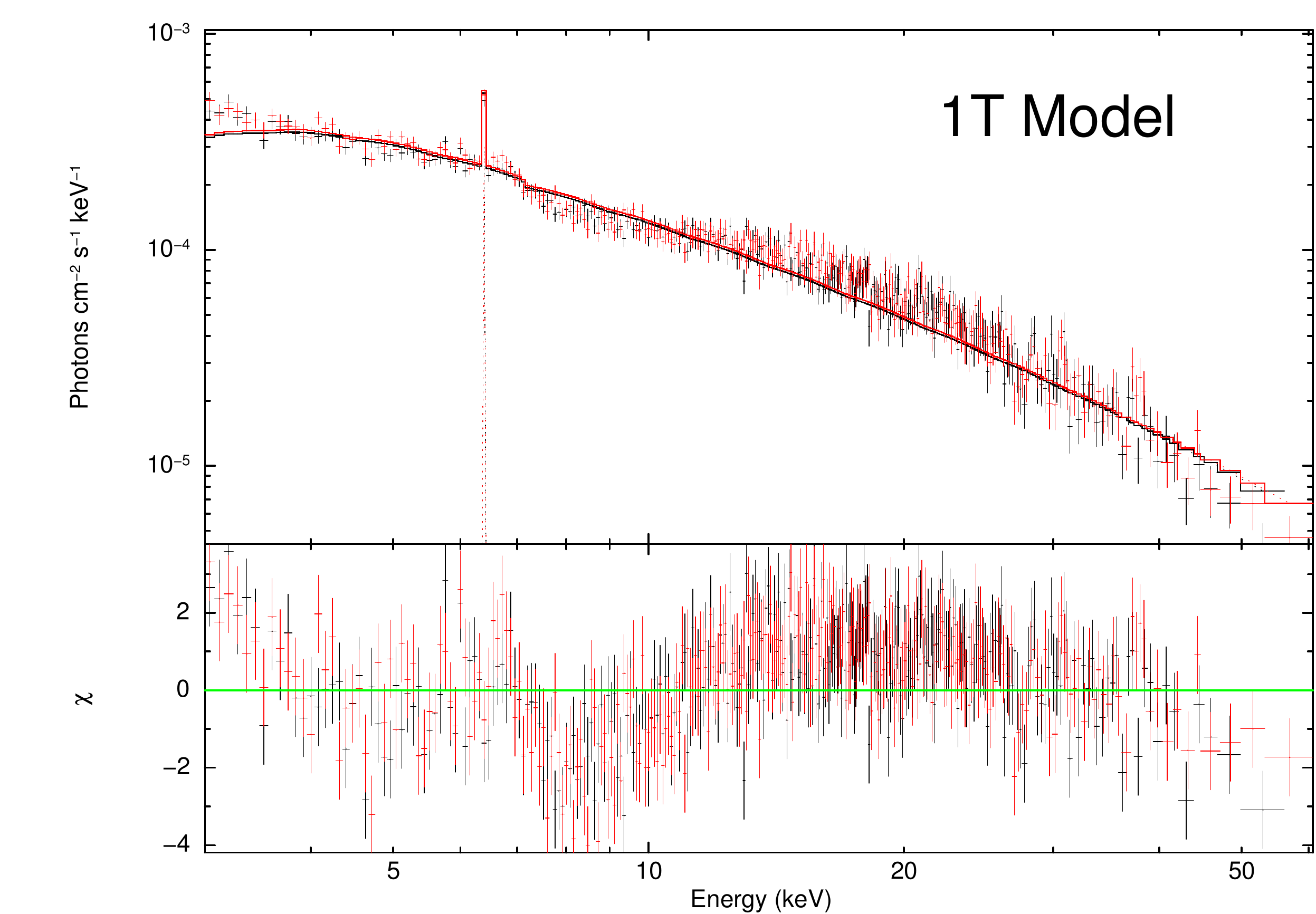,height=0.5\linewidth, angle=270}
\psfig{figure=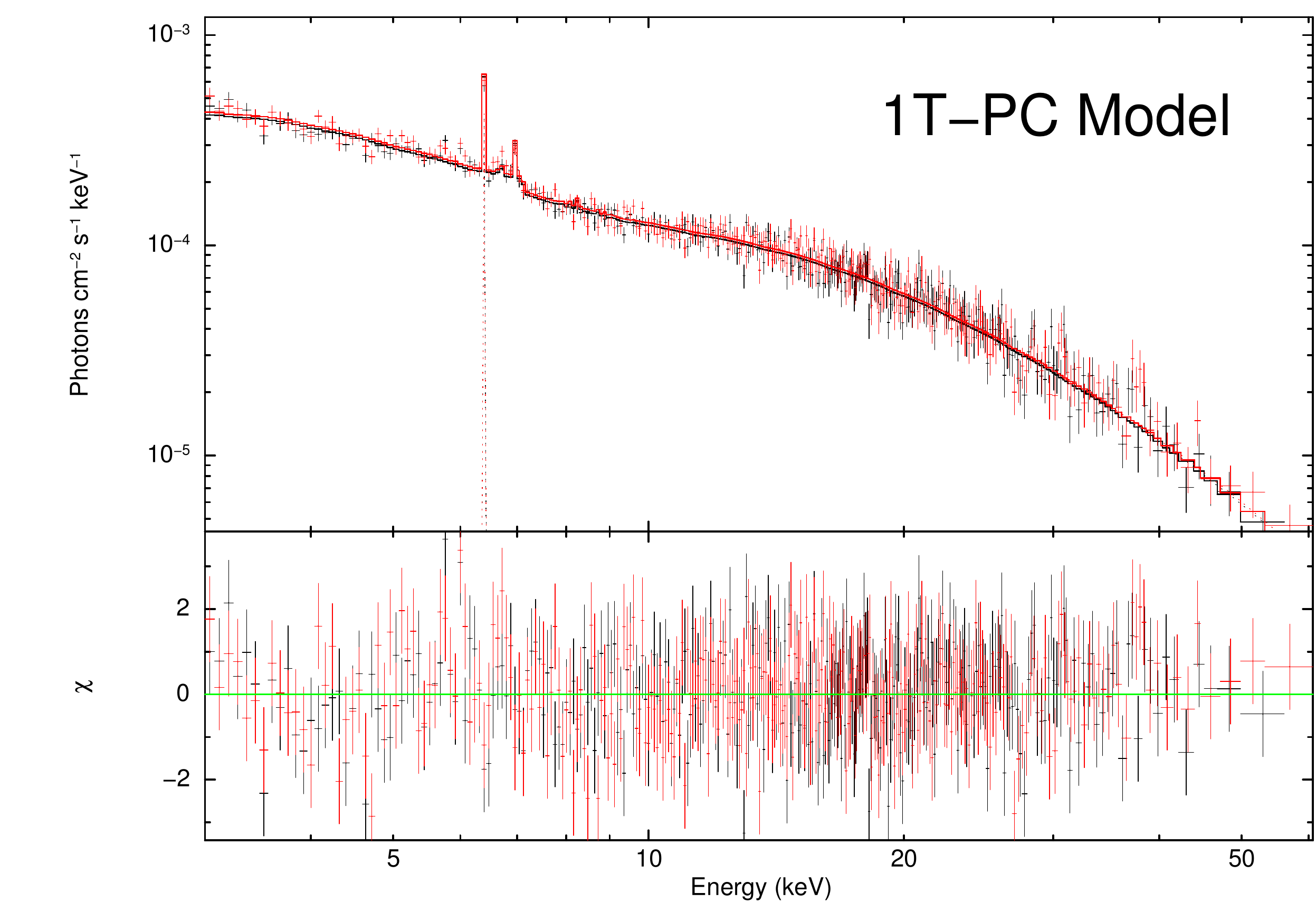,height=0.5\linewidth, angle=270}
}
\mbox{
\psfig{figure=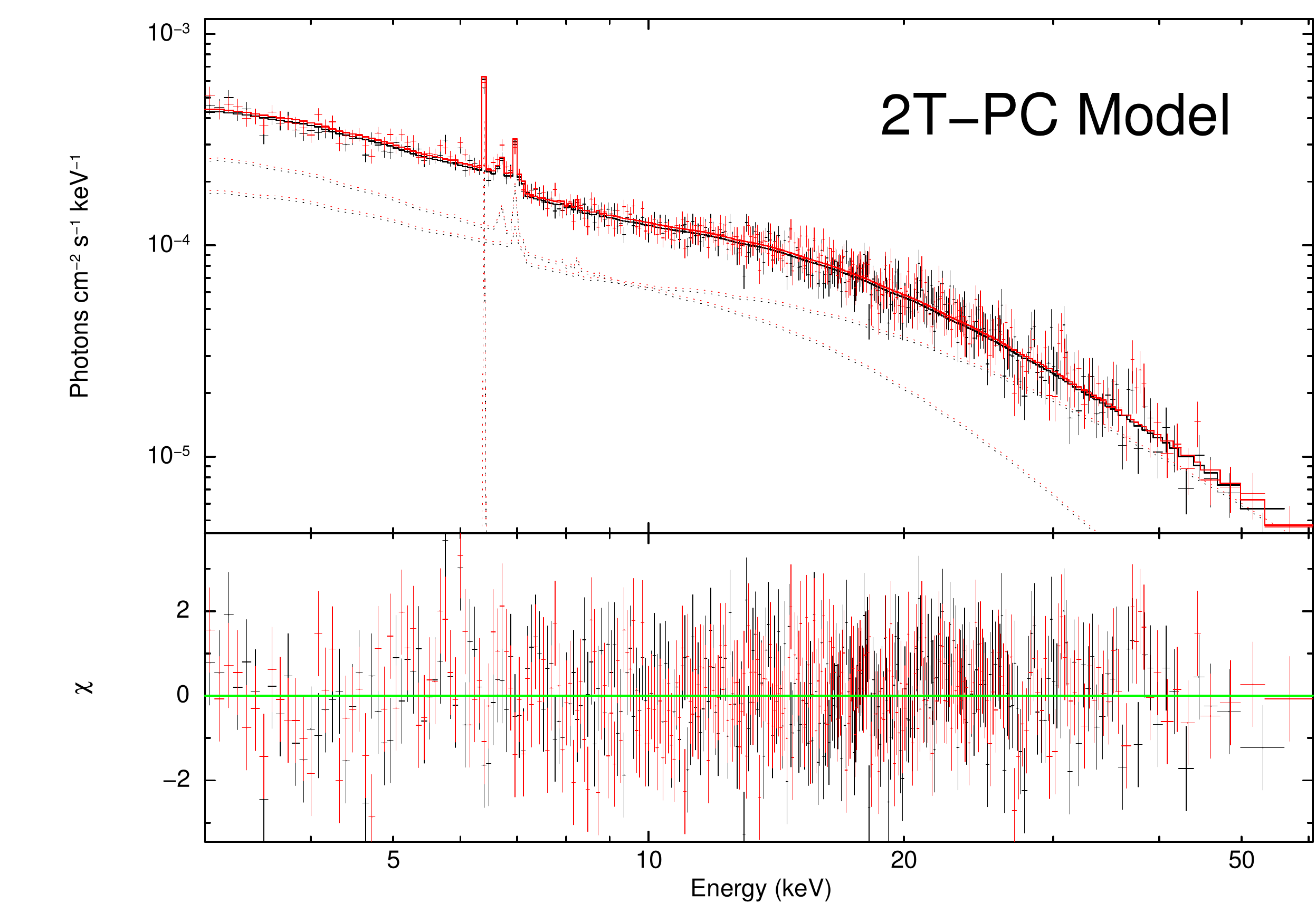,height=0.5\linewidth, angle=270}
\psfig{figure=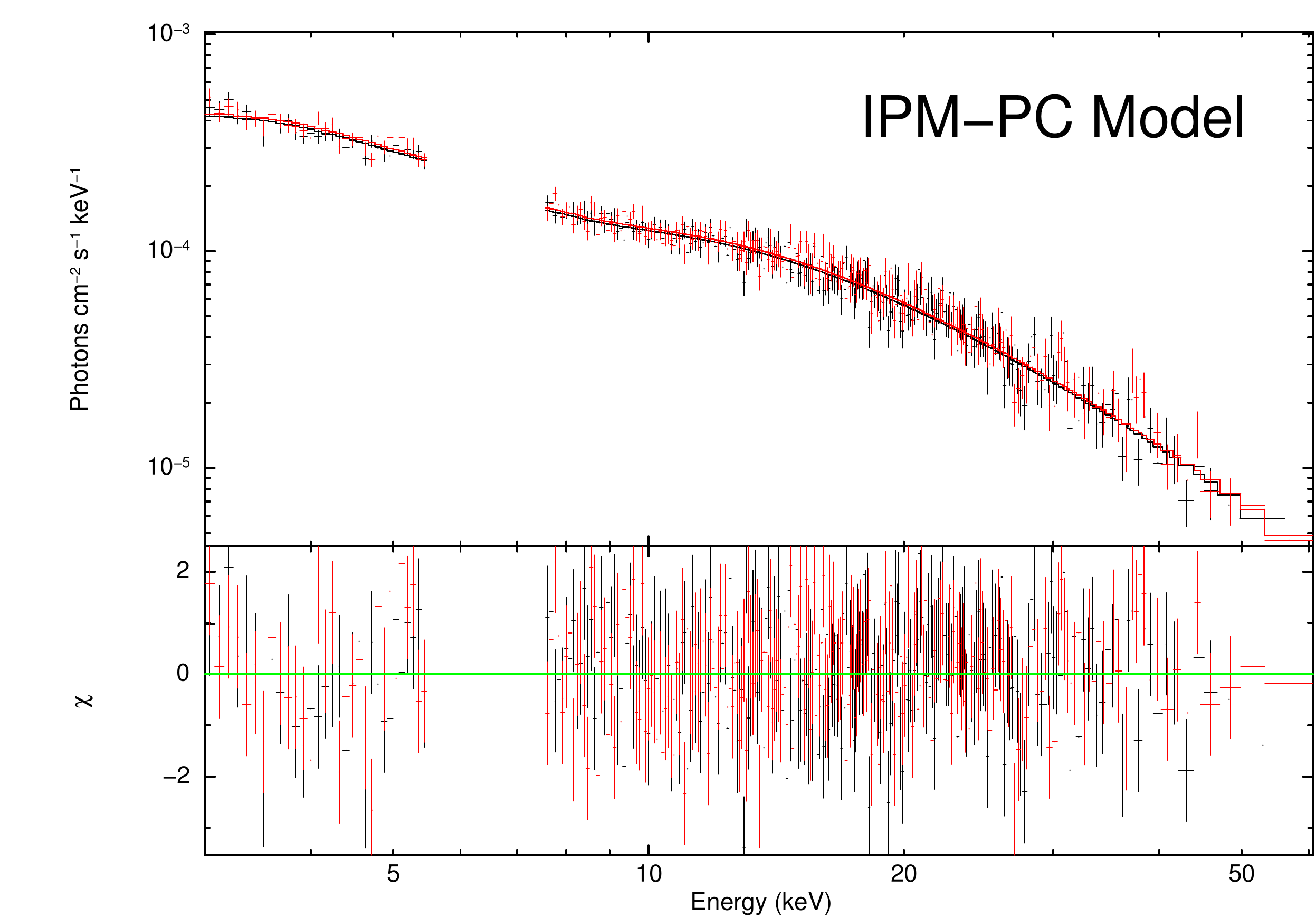,height=0.5\linewidth, angle=270}
}

\caption{\nustar\ 3-50~keV unfolded spectra of \igrsrc\ (Black: FPMA, Red: FPMB) fit with 1T, 2T and IPM models.
A Compton reflection hump from reflection of X-rays off the WD surface is clearly visible in the residuals above 20~keV (top left panel). The spectral fits with the reflection model are not shown here since they exhibit similar residuals
as in the models with partial covering (see Table~\ref{tab:igr_fit} for the fitting results with the reflection model).}
\label{fig:igr}
\end{figure*}

We present figures \ref{fig:tvcol}, \ref{fig:igr} and tables \ref{tab:tvcol_fit}, \ref{tab:igr_fit} for the spectral fitting results for \tvcol\ and \igrsrc. The relevant texts can be found in \S\ref{sec:tvcol_results} and \S\ref{sec:igr_results}.

\begin{deluxetable*}{lcccccccccc}
\tablecaption{{\it NuSTAR} spectral fitting results of \tvcol}
\tablewidth{0pt}
\tablecolumns{11}
\tablehead{ \colhead{Parameters}  & \colhead{1T} & \colhead{1T-PC} & \colhead{1T-PC} & \colhead{1T-RFL} & \colhead{2T} & \colhead{2T-PC\tablenotemark{a}} & \colhead{IPM} & \colhead{IPM} & \colhead{IPM-PC} & \colhead{IPM-RFL} }
\startdata
Bandpass [keV]\tablenotemark{b} & 3-50 & 3-10 & 3-50 & 3-50 & 3-50 & 3-50 & 3-50 & 15-50 & 3-50 & 3-50 \\
$N^{f}_{\rm H}$~[$10^{22}$\,cm$^{-2}$] & $4.6\pm0.4$ & $3.1^{+1.7}_{-2.7}$ & $5.8\pm0.3$ & $4.1^{+0.4}_{-0.5}$ & $10.4^{+1.8}_{-1.6}$ & $0.03^{+4.05}_{-0.02}$ & $8.2\pm0.3$ & --- & $2.6^{+1.7}_{-2.6}$ & $7.7\pm0.6$  \\
$N^{\rm pc}_{\rm H}$~[$10^{23}$\,cm$^{-2}$] & --- & $7.0^{+2.5}_{-2.3}$ & $40\pm4$ & --- & --- & $4.2^{+1.5}_{-1.1}$ & --- & --- & $4.6^{+1.5}_{-1.1}$ & --- \\
$f^{\rm pc}$  & --- & $0.48\pm0.07$ & $0.31^{+0.02}_{-0.03}$ & --- & --- & $0.43^{+0.07}_{-0.6}$ & --- & --- & $0.4\pm0.1$ & --- \\
$kT_{\rm low}$~[keV] & --- & $9.8\pm0.9$ & --- & --- & $1.2^{+0.3}_{-0.2}$ & $2.3^{+0.8}_{-0.5}$ & --- & --- & --- & ---  \\
$Z$\tablenotemark{c} & $1.3\pm0.1$ & $0.36^{+0.08}_{-0.07}$ & $0.93^{+0.11}_{-0.09}$ & $0.9\pm0.1$ & $1.2^{+1.3}_{-1.3}$ & $0.7\pm0.1$ & --- & --- & --- & $0.5$ \\
$N_{\rm low}$~[$10^{-2}$cm$^{-3}$] & --- & $7.2\pm1.0$ & --- & --- & $5.1^{+3.9}_{-2.3}$ & $4.2^{+2.0}_{-1.4}$ & --- & --- & --- & ---  \\
$kT_{\rm high}$~[keV] & $20.7\pm0.4$ & --- & $15.6\pm0.4$ & $16.7^{+0.7}_{-1.9}$ & $21.1\pm0.7$ & $19.7^{+0.9}_{-0.8}$ & --- & --- & --- & --- \\
$N_{\rm high}$~[$10^{-2}$cm$^{-3}$] & $3.6\pm0.1$ & --- & $5.8^{+0.5}_{-0.4}$ & $3.5^{+0.3}_{-0.4}$ & $3.7\pm0.1$ & $4.2\pm0.2$ & --- & --- & --- & ---  \\
$M_{\rm WD}$~[$M_\odot$] & --- & --- & --- & --- & --- & --- & $0.91\pm0.01$ & $0.77\pm0.03$ & $0.80^{+0.01}_{-0.02}$ & $0.78\pm0.03$ \\
$N_{\rm IPM}$~[$10^{-11}$] & --- & --- & --- & --- & --- & --- & $1.8^{+0.3}_{-0.1}$ & $3.0^{+0.4}_{-0.3}$ & $2.7^{+0.1}_{-0.3}$ & $2.2^{+0.4}_{-0.2}$ \\
$\Omega/2\pi$ & --- & --- & --- & $0.6^{+0.4}_{-0.2}$ & --- & --- & --- & --- & --- & $0.8\pm0.2$ \\
$Z_{\rm Fe}$ & --- & --- & --- & $0.9$ & --- & --- & --- & --- & --- & $0.5$ \\
$\cos(\theta)$ & --- & --- & --- & $0.3$ & --- & --- & --- & --- & --- & $0.3$ \\
%Fe line energy [keV] & $6.4$ & $6.4$ & $6.4$ & $6.4$ & $6.4$ & --- & --- & --- & --- \\                                                                                                                                                                    
Fe K$\alpha$ line flux\tablenotemark{d} & $1.1\pm0.1$ & $0.9\pm0.1$ & $1.0\pm0.1$ & $1.0\pm0.1$ & $0.9\pm0.1$ & $0.8\pm0.1$ & --- & --- & --- & ---  \\
Fe K$\alpha$ EW~[eV] & $182\pm14$ & $88\pm15$ & $158\pm14$ & $164\pm15$ & $123\pm14$ & $91\pm14$ & --- & --- & --- & --- \\
Flux~(2-10 keV)\tablenotemark{e} & $4.3$ & $4.3$ & $4.3$ & $4.3$ & $4.2$ & $4.4$ & $4.0$ & $7.1$ & $4.2$ & $4.0$ \\
Flux~(3-50 keV)\tablenotemark{e} & $9.6$ & $7.6$ & $9.5$ & $9.4$ & $9.7$ & $9.6$ & $9.6$ & $11.5$ & $9.5$ & $9.5$ \\
$\chi_\nu^2$~(dof) & 1.33~(596) & 1.07~(169) & 1.22~(594) & 1.26~(594) & 1.18~(594) & 1.10~(593)  & 1.33~(546) & 0.96~(299) & 1.14~(544) & 1.22~(544)
\enddata
\tablecomments {See Table~\ref{tab:models} for the model definitions.}
\tablenotetext{a}{Both APECs are partially covered.}
\tablenotetext{b}{For all IPM models we ignored the iron line complex (5.5-7.5~keV data bins).}
\tablenotetext{c}{Abundance $Z$ for low and high $kT$ components are linked.}
\tablenotetext{d}{The flux unit is $10^{-4}$ph~cm$^{-1}$~s$^{-1}$. Note that we fixed the line energy to 6.4 keV. }
\tablenotetext{e}{The flux unit is $10^{-11}$ \eflux.}
\label{tab:tvcol_fit}
\end{deluxetable*}

\begin{deluxetable*}{lccccccccc}
\tablecaption{{\it NuSTAR} spectral fitting results of \igrsrc}
\tablewidth{0pt}
\tablecolumns{9}
\tablehead{\colhead{Parameters}  & \colhead{1T} & \colhead{1T-PC} & \colhead{1T-PC} & \colhead{1T-RFL} & \colhead{2T-PC} & \colhead{IPM} & \colhead{IPM-PC} & \colhead{IPM-RFL} & \colhead{IPM-PC-RFL}}
\startdata
Bandpass [keV]\tablenotemark{a} & 3-50 & 3-10 & 3-50 & 3-50 & 3-50 & 15-50 & 3-50 & 3-50 & 3-50 \\
$N^{f}_{\rm H} [10^{22}$\,cm$^{-2}$] & $7.8^{+0.6}_{-0.5}$ & $2.9^{+2.4}_{-2.5}$ &  $4.9\pm0.5$ & $3.0\pm0.6$ & $4.9$ & --- & $6.3\pm0.6$ & $3.0^{+1.0}_{-1.1}$ & $3.0^{+1.0}_{-1.1}$ \\
$N^{\rm pc}_{\rm H} [10^{23}$\,cm$^{-2}$] & --- & $11\pm3 $ & $28\pm2$ & --- & $26\pm2$ & --- & $22\pm1$ & --- & $11^{+8}_{-2}$ \\
$f^{\rm pc}$  & --- & $0.66\pm0.08$ & $0.56^{+0.01}_{-0.02}$ & --- & $0.59\pm0.03$ & --- & $0.63\pm0.01$ & --- & $0.43^{+0.08}_{-0.11}$ \\
$kT_{\rm low}$ [keV] & --- & --- & --- & --- & $14^{+6}_{-10}$ & --- & --- & --- & --- \\
$Z$\tablenotemark{b} & $0.3$ & $0.3$ & $0.3\pm0.1$ & $0.3^{+0.2}_{-0.1}$ & $0.3\pm0.1$ & --- & --- & $0.4^{+0.2}_{-0.1}$ & $0.3^{+0.3}_{-0.2}$ \\
$N_{\rm low}$ [$10^{-2}$cm$^{-3}$] & --- & --- & --- & --- & --- & --- & $2.3\pm1.0$ & --- \\
$kT_{\rm high}$ [keV] & $64.0_{-0.5}$ & $10.2_{-1.1}^{+3.5}$ & $26\pm1$ & $41^{+3}_{-2}$ & $43\pm12$ & --- & --- & --- & --- \\
$N_{\rm high}$ [$10^{-2}$cm$^{-3}$] & $1.94^{+0.02}_{-0.05}$ & $3.79\pm0.01$ & $3.5\pm0.3$ & $1.3^{+0.4}_{-0.2}$ & $1.7^{+1.0}_{-0.5}$ & --- & --- & --- & --- \\
$M_{\rm WD}$ [$M_\odot$] & --- & --- &  --- & --- & --- & $1.16\pm0.05$ & $0.98\pm0.03$ & $1.34\pm0.02$ & $1.05^{+0.03}_{-0.02}$ \\
$N_{\rm IPM}$ [$10^{-11}$] & --- & --- & --- & --- & --- & $0.8\pm0.1$ & $1.5\pm0.3$ & $0.2\pm0.5$ & $0.6^{+0.3}_{-0.2}$ \\
$\Omega/2\pi$ & --- & --- & --- & $1.0$ & --- & --- & --- & $1.0$ & $0.8^{+0.4}_{-0.3}$ \\
$Z_{\rm Fe}$ & --- & --- & --- & $0.3$ & --- & --- & --- & $0.4$ & $0.3$ \\
$\cos(\theta)$ & --- & --- & --- & $0.95_{-0.06}$ & --- & --- & --- & $0.95_{-0.01}$ & $0.95_{-0.5}$ \\
Fe K$\alpha$ line flux\tablenotemark{c} & $2.5\pm0.5$ & $1.5_{-0.7}^{+0.9}$& $3.5\pm0.5$ & $2.2\pm0.5$ & $7.6\pm1.1$ & --- & --- & --- & --- \\
Fe K$\alpha$ EW [eV] & $97\pm20$ & $58\pm19$ & $146\pm22$ & $89\pm24$ & $134^{+12}_{-23}$ & --- & --- & --- & --- \\
Flux (2-10 keV)\tablenotemark{d} & $1.7$ & $1.8$ & $1.7$ & $1.7$ & $1.7$ & $3.9$ & $1.6$ & $1.7$ & $1.7$ \\
Flux (3-50 keV)\tablenotemark{d} & $6.6$ & $3.8$ & $6.7$ & $6.8$ & $6.7$ & $8.5$ & $6.7$ & $6.8$ & $6.7$ \\
$\chi_\nu^2$ (dof) & 1.90 (598) & 1.10 (168) & 1.05 (595) & 1.03 (596) & 1.04 (594) & 1.02 (300) & 1.00 (545) & 1.04 (544) & 0.97 (543)
\enddata
\tablecomments {See Table~\ref{tab:models} for the model definitions.}
\tablenotetext{a}{For all IPM models we ignored the iron line complex (5.5-7.5~keV data bins).}
\tablenotetext{b}{Abundance $Z$ for low and high $kT$ components are linked.}
\tablenotetext{c}{The flux unit is $10^{-5}$ph~cm$^{-1}$~s$^{-1}$. Note that we fixed the line energy to 6.4~keV.}
\tablenotetext{d}{The flux unit is $10^{-11}$ \eflux.}
\label{tab:igr_fit}
\end{deluxetable*}

%\bibliography{NuSTAR_IP}

\end{document}